\documentclass[useAMS,usenatbib]{mnras}
\usepackage{graphicx,amsmath}
\usepackage{amssymb}
\usepackage{natbib}
\usepackage{color}
\title[Constraining QSO SED]{Constraints on QSO emissivity using H~{\sc i} and He~{\sc ii} Lyman alpha forest}
\author[Vikram Khaire]
{
\parbox{\textwidth}{ 
Vikram Khaire \thanks{E-mail:kvikram@ncra.tifr.res.in}  
} 
\vspace*{10pt}\\ 
National Centre for Radio Astrophysics, 
Tata Institute of Fundamental Research, Pune 411007, India\\
 }   

\begin{document}

\defcitealias{Khaire15ebl}{KS15b}
\defcitealias{Khaire15puc}{KS15a}
\defcitealias{HM12}{HM12}
\defcitealias{Madau15}{MH15}

\date{}

\pagerange{\pageref{firstpage}--\pageref{lastpage}} \pubyear{2017}
\maketitle

\label{firstpage}

\vspace{20 mm}

\newcommand{\tHI}{$\tau_{\rm \alpha}^{\rm H\,I}$}
\newcommand{\tHeII}{$\tau_{\rm \alpha}^{\rm He\,II}$}
\begin{abstract} 
The spectrum of cosmic ultraviolet background radiation at He~{\sc ii}
ionizing energies ($\rm E \geq 4\,Ryd$) is important to study the He~{\sc ii} 
reionization, thermal history of the intergalactic medium (IGM) and metal 
lines observed in QSO absorption spectra. It is determined by the 
emissivity of QSOs at $\rm E \geq 4\,Ryd$ obtained from their observed 
luminosity functions and the mean spectral energy distribution (SED). The 
SED is approximated as a power-law at energies $\rm E \geq 1$ Ryd, 
$f_{E}\propto \rm E ^{\alpha}$, where the existing observations constrain 
the power-law index $\alpha$ only up to $\sim$ 2.3 Ryd. Here, we constrain 
$\alpha$ for $\rm E \geq 4\,Ryd$ using recently measured He~{\sc ii} Lyman-$\alpha$ 
effective optical depths (\tHeII), H~{\sc i} photoionization rates
and updated H~{\sc i} distribution in the IGM. 
We find that $-1.6>\alpha>-2$ is required to reproduce the \tHeII~measurements 
when we use QSO emissivity obtained from their luminosity function using 
optical surveys. We also find that the models where QSOs can alone reionize 
H~{\sc i} can not reproduce the \tHeII~measurements. These models need 
modifications, such as a break in mean QSO SED at energies greater than 4 Ryd.
Even after such modifications the predicted He~{\sc ii} reionization   history, 
showing that the He~{\sc ii} is highly ionized even at $z\sim5$, 
is significantly different from the standard models. 
Therefore, the thermal history of the IGM will be crucial to distinguish these models.
We also provide the He~{\sc ii} photoionization rates obtained from binned 
\tHeII~measurements. 
\end{abstract}


\section{Introduction}
The observed ionization state of the intergalactic medium (IGM) at $z\le6$ 
\citep{Gunn65, Fan06, Becker13} is maintained by cosmic ultraviolet background 
(UVB) radiation emanating from Quasi-stellar Objects (QSOs) and galaxies 
\citep[][]{Miralda90, Shapiro94, HM96, Shull99}. Apart from being the main driver 
of the hydrogen and helium reionization, the UVB maintains the ionization state 
of metals in the IGM and in the circum-galactic environments of galaxies. 
Therefore, the spectrum of UVB is important to 
study the cosmic metal mass density and the metal enrichment of the IGM 
\citep[see for e.g.;][]{Songaila96, Songaila01, Carswell02, Bergeron02, 
Simcoe04, Shull14, Peeples14, Hussain17} by relating 
the observed ionic abundances to metal abundances.
%
%
%
\begin{table*}
\caption{Measurements of the power-law index $\alpha$ (where $f_{\nu}\propto \nu ^{\alpha}$)}
\def\arraystretch{1.5}
\begin{center}
\begin{tabular}{ l c c c c c}
              
\hline
 Reference          & $\alpha$             &  $\lambda_{\rm rest}$-Range  & N$_{\rm QSOs}$   & $z$-Range       &  Survey  \\
 (1)          & (2)             &  (3)& (4)& (5)      &  (6)  \\
\hline                                                                     
 \citet{Telfer02}    &  -1.57$\pm$0.17     &  $500-1200$\AA               &  77     &   $>0.33$    &  HST/FOS (radio-quite sample)  \\
 \citet{Telfer02}    &  -1.96$\pm$0.12     &  $500-1200$\AA               & 107    &    $>0.33$    &   HST/FOS (radio-loud sample)\\
 \citet{Scott04}     &  -0.56$\pm$0.38     &  $630-1150$\AA               &  85     &  $<0.67$     &  FUSE       \\
 \citet{Shull12}     &   -1.41$\pm$0.21    &  $550-1000$\AA               &  15     &  $0.45-1.44$ &   HST/COS    \\
 \citet{Stevans14}   &   -1.41$\pm$0.15    &  $500-1000$\AA               & 159     &   $<1.476$   &   HST/COS \\
 \citet{Lusso15}     &   -1.70$\pm$0.61    &   $600-912$\AA                &  53      &  $2.3-2.6$  &  HST/WFC3 \\
 \citet{Tilton16}    &  -0.72$\pm$0.26     &  $425-850$\AA                &  20     &  $1.0-2.1$   &   HST/COS \\
\hline
\end{tabular}
\end{center}
\begin{flushleft}
\footnotesize{{\bf Notes:}}\\
\footnotesize{Column (1) gives references. Column (2) provides the measurements 
of $\alpha$ with the quoted 1-$\sigma$ errors measured for the rest wavelength 
($\lambda_{\rm rest}$) range as given in column (3). Column (4) shows the number 
of QSOs (N$_{\rm QSOs}$) used to obtain the composite spectrum having emission 
redshift as given in column (5). Column (6) provides information about survey, 
i.e the instrument, telescope and sample characteristics, where FOS stands for the 
Faint Object Spectrograph and WFC3 stands for the Wide Field Camera 3 on board HST.}
\end{flushleft}
\label{t2}
\end{table*}
%
%

Spectrum of the UVB depends on the spectral energy distribution (SED) of the 
sources that are contributing to it, mainly QSOs and star-forming galaxies. 
If we divide the UVB naively into hydrogen ionizing part (1 Ryd$<\rm E<$4 Ryd) 
and helium ionizing part ($\rm E \geq 4 $ Ryd), the former is contributed by both 
galaxies and QSOs but latter is predominantly contributed by only QSOs. The relative 
contribution by QSOs and galaxies to the hydrogen ionizing part of the UVB depends 
on average escape fraction ($f_{\rm esc}$), a parameter that quantifies 
the amount of hydrogen ionizing photons escaping from galaxies.
The $f_{\rm esc}(z)$ can be obtained using the measurements 
of hydrogen photoionization rates ($\Gamma_{\rm HI}$) 
for a given QSO emissivity and star formation history of galaxies \citep[see][]{Inoue06, Khaire16}. 
On the other hand, for the measured 
$\Gamma_{\rm HI}(z)$ and the H~{\sc i} distribution in the IGM, the helium 
ionizing part of the UVB depends only on the QSO emissivity at $\rm E \geq 4$ Ryd.
This emissivity is estimated through QSO luminosity functions and the mean SED of QSOs. 
The SED is usually approximated as a power-law, 
$f_{\nu}\propto \nu ^{\alpha}$ at $\rm E \geq 1$ Ryd ($\lambda \leq 912$\AA) from the observed 
composite QSO spectra \citep[][]{Zheng97, Telfer02, Scott04, Stevans14, Lusso15}.
Although the existing observations have probed mean QSO SED only up to 
E$\sim$2.3 Ryd ($\lambda\sim 400$\AA), it is usually extrapolated up to 35 Ryd 
($\lambda\sim 25$\AA) to calculate the He~{\sc ii} ionizing emissivity and the UVB. 
The reported values of the power-law index $\alpha$ show large variation from 
 $-0.56$ to $-1.96$. Moreover, the number of QSOs where SED at high-energies can 
be directly probed is very small \citep[see for e.g.,][]{Tilton16}. The existing 
measurements of $\alpha$ over the last two decades are summarized in Table~\ref{t2}. 
Using different $\alpha$ in UVB models gives significantly different UVB 
spectrum especially for $\rm E \geq 4$ Ryd. Also, the He~{\sc ii} ionizing emissivities 
obtained using different $\alpha$ provide different histories of the He~{\sc ii} 
reionization. Like hydrogen ionizing part of the UVB, we need measurements of 
He~{\sc ii} photoionization rates ($\Gamma_{\rm HeII}$) that can be used to 
constrain the He~{\sc ii} ionizing emissivity. The accurate estimate of UVB 
spectrum, especially at ${\rm E} \geq 4$ Ryd ($\lambda \leq 228$\AA), is important for studying 
the ionization mechanism for high ionization systems such as O~{\sc vi} 
\citep[see for e.g][]{Danforth05, Tripp08, Muzahid12o6, Pachat16} and 
Ne~{\sc viii} \citep[see for e.g.;][]{Savage05, Savage11, Narayanan12, 
Meiring13, Hussain15, Hussain17} which are believed to trace the warm-hot 
phase of the IGM. It is also important for studying the thermal history of the 
IGM \citep{Lidz10, Bolton10, Becker11t, Bolton12, Khrykin17} and the process of 
He~{\sc ii} reionization \citep{FG09, McQuinn09, Compostella13, Plante16}. 
The above mentioned importance of $\alpha$ and the issues with its measurements motivate us 
to theoretically constrain $\alpha$ at ${\rm E \ge 4}$~Ryd. For that we use the observations of
H~{\sc i} and He~{\sc ii} Lyman-$\alpha$ forest.

The He~{\sc ii} Lyman-$\alpha$ forest has been observed for few QSOs at 
$z>2.5$ with UV spectrographs on space telescopes such as Far Ultraviolet 
Spectroscopic Explorer \citep[FUSE;][]{Kriss01, Shull04, Fechner06} and Cosmic 
Origin Spectrograph (COS) on-board Hubble Space Telescopes 
\citep[HST;][]{Syphers11, Worseck16}. With such observations the Lyman-$\alpha$ 
effective optical depths of He~{\sc ii} \citep[\tHeII;][]{Shull10, Syphers13, 
Worseck11} and the ratio of He~{\sc ii} to H~{\sc i} in the IGM absorbers 
\citep{Zheng04, Muzahid11, McQuinn14} have been measured. The recent 
measurements of \tHeII~by \citet{Worseck16} at $2.3<z<3.5$ can be used to 
constrain the He~{\sc ii} ionizing emissivity and the properties of QSO SED such 
as the spectral index $\alpha$. This is what we explore in our analysis.

For a given QSO emissivity at $1$ Ryd and a mean SED of QSOs, using our
cosmological radiative transfer code \citep{Khaire13, Khaire15ebl, Khaire15puc}, 
we estimate the He~{\sc ii} ionizing UVB, photoionization rates of He~{\sc ii} 
and \tHeII.  We also calculate the corresponding He~{\sc ii} reionization history.
By comparing these values with the \tHeII~measurements, we constrain the mean SED of 
QSOs. We use two models of QSO emissivity, one obtained from the compilation of 
optically selected QSOs \citep{Khaire15puc} and the other where QSOs 
can alone reionize H~{\sc i} when extrapolated to $z>6$ \citep{Madau15, Khaire16}. 
The latter uses the QSO luminosity function of \citet{Giallongo15} that claimed to 
detect large number density of low luminosity QSOs at $z>4$. Using \tHeII~and 
$\Gamma_{\rm HI}$ measurements we also estimate the $\Gamma_{\rm HeII}$ values 
that depends only on the H~{\sc i} distribution of the IGM and independent of 
the UVB models.

The paper is organized as follows. In section~\ref{sec2}, we discuss the basic 
theory to calculate \tHeII~using  H~{\sc i} distribution of the IGM and 
$\Gamma_{\rm HeII}$ using \tHeII~measurements. In Section~\ref{sec3}, 
we explain the basic theory and assumptions to calculate the He~{\sc ii} 
ionizing emissivity, the UVB and the He~{\sc ii} reionization history.  
In Section~\ref{sec4}, we discuss our results for different models of 
QSO emissivity and uncertainties. We present the summary in section~\ref{sec5}. 
Throughout this paper we use cosmology parameters $\Omega_{\Lambda}=0.7$, 
$\Omega_{m}=0.3$ and $H_0=70$ km s$^{-1}$ Mpc$^{-1}$ consistent with that 
from \citet{Planck16}.
%
%
\section{He~{\sc ii} optical depths and photoionization rates}\label{sec2}
\subsection{Basic theory: Lyman-$\alpha$ effective optical depths}\label{sec2.1}
The Lyman-$\alpha$ effective optical depth for H~{\sc i} (\tHI) and 
He~{\sc ii} (\tHeII) at redshift $z$ is obtained by \citep{Paresce,Madau94},
%
\begin{equation}\label{tau_he}
\tau_{\alpha}^{x}(z)=\frac{1+z}{\lambda^{^x}_{\alpha}}
\int_{N^{\rm min}_{\rm HI}}^{\infty}dN_{\rm HI}\,
\frac{\partial^{2}N}{\partial N_{\rm HI}\partial z}\,W_{\rm n}^{x}\,\,.
\end{equation}
%
Here, $x$ denotes the species H~{\sc i} or He~{\sc ii}, 
$\lambda^{^x}_{\alpha}$ is the rest-frame Lyman-$\alpha$ line wavelength of species $x$ 
(i.e, 1215.67\AA~for H~{\sc i} and 303.78\AA~for He~{\sc ii}), 
$N_{\rm H\,I}^{\rm min}$ is the minimum column density of H~{\sc i} used in the 
integral and $\partial^{2}N/\partial N_{\rm HI}\partial z=f(N_{\rm HI},z)$ 
is the column density distribution of 
H~{\sc i}. Here, $W_{\rm n}^{x}$ is the equivalent width 
of the Lyman-$\alpha$ line expressed in wavelength units for 
species $x$ as given by,
%
\begin{equation}\label{wn}
W_{\rm n}^{x}=\int_{0}^{\infty}d\lambda\,[1-e^{-y\phi_{x}(\lambda)}]\,\,,
\end{equation}
%
where, $\phi_x(\lambda)$ is the Voigt profile function for species $x$, 
$y=N_{\rm HI}$ when $x$ is  H~{\sc i} and $y=\eta \times N_{\rm HI}$ when
$x$ is He~{\sc ii} where $\eta=N_{\rm He II}/N_{\rm HI}$.

The calculation of \tHI~depends on  the observed $f(N_{\rm HI},z)$. 
In the absence of the column density distribution of He~{\sc ii}, the 
calculation of \tHeII~relies on the the estimate of the parameter $\eta$. 
The $\eta$ determines the amount of $N_{\rm He II}$ in intergalactic absorber 
having H~{\sc i} column density $N_{\rm HI}$. It is estimated under the 
assumption that the IGM is in photoionization equilibrium maintained by 
the UVB. 
The $\eta$ is independent of $N_{\rm HI}$ for the absorbers that are 
optically thin to He~{\sc ii} ionizing radiation 
($N_{\rm HeII}\la 10^{16.8}{\rm cm^{-2}}$; obtained for continuum optical 
depth $ \la 0.1$), called as $\eta_{\rm thin}$. 
The parameter $\eta_{\rm thin}$ is obtained from the relation,
%
\begin{equation}\label{eta_thin}
\eta_{\rm thin}(z)=\frac{n_{\rm He}}{n_{\rm H}}
\frac{\alpha^A_{\rm HeII}(\rm T)}{\alpha^A_{\rm HI}(\rm T)}
\frac{\Gamma_{\rm HI}(z)}{\Gamma_{\rm HeII}(z)}.
\end{equation}
%
Here, $\alpha^A_{x}(\rm T)$ and $\Gamma_x$ 
are the case A recombination rate coefficient (that depends on the gas temperature T) and 
the photoionization rate for species $x$, respectively,
whilst $n_{\rm H}$ and $n_{\rm He}$ are the number density of total hydrogen and 
helium in the IGM, respectively. The ratio 
$n_{\rm He}/n_{\rm H}=y_p/(4-4y_p)$ where $y_p$ is the primordial mass 
fraction of helium. Using $y_p=0.25$ from \citet{Planck16} and 
the expressions for recombination rate coefficients\footnote{ The case A 
recombination rate coefficients for H~{\sc i} and He~{\sc ii}  in units of cm$^3$ s$^{-1}$ are given by
$\alpha^A_{\rm HI}=2.51 \times 10^{-13} T^{-0.76}_{4.3}$ 
and $\alpha^A_{\rm HeII}=1.36 \times 10^{-12} T^{-0.70}_{4.3}$ where $T=10^{4.3}T_{4.3}$K.},
Eq.~\ref{eta_thin} can be approximated as,  
%
\begin{equation}\label{eta_thin_approx}
\eta_{\rm thin}(z)=0.45 \Big(\frac{\rm T}{10^{4.3}{\rm K}}\Big)^{0.06}
\frac{\Gamma_{\rm HI}(z)}{\Gamma_{\rm HeII}(z)}.
\end{equation}
%
The above equation shows that $\eta_{\rm thin}$ weakly depends on the 
temperature and it is mainly decided by the ratio of $\Gamma_{\rm HI}$ 
to $\Gamma_{\rm HeII}$. Under photoionization equilibrium, $\eta$ at all 
$N_{\rm HI}$ obtained from radiative transfer simulations can be 
approximated by the following quadratic equation \citep{Fardal98, FG09, HM12},
%
\begin{equation}\label{quad}
\begin{aligned}
\frac{n_{\rm He}}{4n_{\rm H}}\,
\frac{\Gamma_{\rm HI}}{n_e\alpha^A_{\rm HI}(\rm T)}\,
\frac{\sigma_{912}N_{\rm HI}}{(1+{\rm A}\sigma_{912}N_{\rm HI})}
=\sigma_{228}N_{\rm HeII}\\
\,+\,\frac{\Gamma_{\rm HeII}}{n_e \alpha^A_{\rm HeII}(\rm T)}\,
\frac{\sigma_{\rm HeII}N_{\rm HeII}}{(1+{\rm B}\sigma_{228} N_{\rm HeII})}.
\end{aligned}
\end{equation}
%
Here, $n_e$ is electron density, $\sigma_{228}$ is photoionization 
cross-section of He~{\sc ii} ($\sigma_{\rm HeII}$) at 228\AA, $\sigma_{912}$ 
is photoionization cross-section of H~{\sc i} ($\sigma_{\rm HI}$) at 912\AA, 
and A and B are the constants obtained by fitting numerical results. 
The above quadratic equation is supplemented by a relation between $n_e$ 
and $N_{\rm HI}$. We take this relation, 
$n_e=1.024\times10^{-6} (N_{\rm HI}\Gamma_{\rm HI})^{(2/3)}{\rm cm^{-3}}$, 
T=20000K, and the values of $A=0.02$ and $B=0.25$ following \citet{HM12}. 
These parameters are obtained for the clouds having plane parallel slab
geometry and fixed line-of-sight length equal to the Jeans length following 
\citet{Schaye01}. With the same set-up, we also verify these values 
using~{\sc cloudy13} \citep{Ferland13}. The $\eta$ obtained by solving 
Eq.~\ref{quad} reduces to $\eta_{\rm thin}$ for optically thin clouds.
Although we use Eq.~\ref{quad} to calculate $\eta$ 
at all $N_{\rm HI}$, the \tHeII~is mainly due to optically thin 
clouds of He~{\sc ii} where $\eta=\eta_{\rm thin}$, therefore, \tHeII~is 
independent of the geometry or the finite size of clouds. 

It is important to set the appropriate $N^{\rm min}_{\rm HI}$ 
in Eq.~\ref{tau_he} since,
$\tau^x_{\alpha}$ depends on it \citep[see also][]{Madau94}. 
It is because, for low column densities 
$W^x_{n} \propto N_x$ and the column density distribution
of H~{\sc i} is a power-law in $N_{\rm HI}$, i.e,
$f(N_{\rm HI}, z)\propto N_{\rm HI}^{-\beta}$ where $\beta$ is a power-law 
index. Using these relations in Eq.~\ref{tau_he} gives  
\tHI~$\propto N^{2-\beta}_{\rm HI}$
and \tHeII~$\propto \eta_{\rm thin}N^{2-\beta}_{\rm HI}$. Therefore, it
is unphysical to extrapolate the power-law $f(N_{\rm HI}, z)$ to smaller 
$N_{\rm HI}$ than what observations suggest.  
We use the parametric form of $f(N_{\rm HI},z)$ from \citet{InoueAK14}.
It reproduces the observed redshift evolution of the \tHI(z) 
\citep[by][]{Fan06, Kirkman07, FG08, Becker13Ly}. \citet{InoueAK14} has used 
$N^{\rm min}_{\rm HI}=10^{12}\,{\rm cm^{-2}}$ and $b$-parameter (mean Doppler 
velocity to estimate the Voigt profile function) of 28 km $s^{-1}$ to calculate 
\tHI~using Eq.~\ref{tau_he}. This corresponds to a 
minimum equivalent width of H~{\sc i} Lyman-$\alpha$ line to be
$W_{n}^{\rm HI}=5.2\times10^{-3} {\rm \AA}$. To calculate 
\tHeII, we use the same $b$-parameter assuming that the Doppler broadening is mostly 
dominated by turbulence and  
$N^{\rm min}_{\rm HI}=(16/\eta_{\rm thin})\times 10^{12}\,{\rm cm^{-2}}$ 
that gives the same minimum equivalent width for He~{\sc ii} as mentioned above 
for H~{\sc i}. In Section~\ref{sec4.3}, we discuss the uncertainty in the 
obtained \tHeII~arising from these assumptions and its effect on the presented results. 

In the following sub-section, we calculate $\eta_{\rm \, thin}$ 
from the \tHeII~measurements and estimate the corresponding 
He~{\sc ii} photoionization rates.

%
%
\begin{figure*}
\centering
\includegraphics[totalheight=0.75\textheight, trim=0.0cm 0.8cm 7.7cm 0.0cm, clip=true, angle=270]{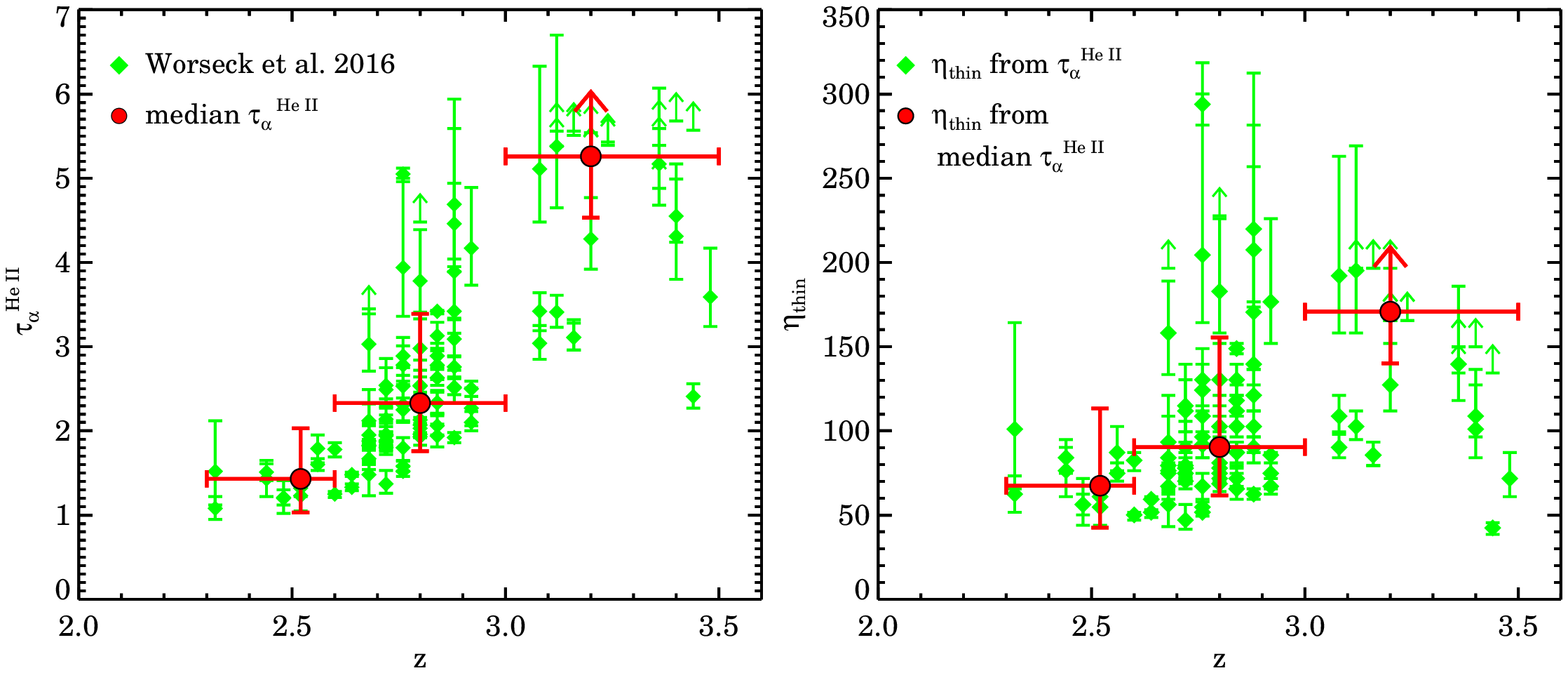}
\caption{Left-hand panel: Measurements of \tHeII($z$) from \citet{Worseck16}. 
The red circles are the median \tHeII~in three redshift bins as given in Table~\ref{t1}. 
Horizontal bars show the sizes of redshift bins. Vertical error-bars on  median 
\tHeII~show 95th percentile values of the distribution of errors in each redshift-bin. 
Right-hand panel: $\eta _{\rm \, thin}=N_{\rm HeII}/N_{\rm HI}$ calculated using the 
\tHeII~data as plotted on the left-hand panel. Red circles show the $\eta_{\rm \, thin}$ 
from the median \tHeII~data in three redshift bins as shown in the the left-hand panel. 
The values are provided in Table~\ref{t1}. }
\label{fig1}
\end{figure*}
%
%
%
\begin{table}
\caption{$\Gamma_{\rm He II}$ and $\lambda_{\rm mfp}$ estimates}
\def\arraystretch{1.7}
\begin{tabular}{ l c c c c }
              
\hline
 Median $z$                                     & 2.52                     & 2.8                       &  3.2   \\
\hline      
 $z$-range                                      & $2.3-2.6$                & $2.6-3.0$                 & $3.0-3.5$ \\
 $^a$median \tHeII~                             & 1.43$^{+0.60}_{-0.40}$   & 2.33$^{+1.06}_{-0.57}$    & 5.26$^{+\infty}_{-0.73}$\\
 $\eta_{\, \rm thin}$                           & 67.4$^{+45.9}_{-24.9}$   & 90.4$^{+65.1}_{-28.7}$    & 170.8$^{+\infty}_{-30.6}$ \\
 $^b$$\Gamma_{\rm HI}$ in $10^{-12}$ s$^{-1}~$  & 1.035$^{+0.37}_{-0.30}$  & 0.86$^{+0.30}_{-0.22}$    & 0.79$^{+0.28}_{-0.19}$     \\
 $\Gamma_{\rm He II}$ in $10^{-15}$ s$^{-1}$    & 6.91$^{+5.31}_{-3.22}$   & 4.28$^{+3.42}_{-1.76}$    & 2.08$^{+1.83}_{-\infty}$  \\
 $ \lambda_{\rm mfp}$ in pMpc                   & 32.9$^{+10.7}_{-9.0}$    & 18.7$^{+5.0}_{-5.4}$      & 7.5$^{+1.0}_{-7.5}$  \\
\hline
\end{tabular}
\footnotesize{{\bf Notes:}}\\
\footnotesize{$^a$Errors on the mean \tHeII~ correspond to 95th percentile 
of the distribution of errors on \tHeII~measurements in the redshift-bin.}
\footnotesize{$^b$$\Gamma_{\rm HI}$ measurements from \citet{Becker13}.}
\hfill
\label{t1}
\end{table}
%
%
%
%
\subsection{He~{\sc ii} photoionization rates}\label{sec2.2}

In Eq.~\ref{tau_he} and \ref{wn}, the value of $\eta_{\, \rm thin}$
can be varied to obtain the desired value of \tHeII. By this method, 
one can obtain the values of  $\eta_{\, \rm thin}$ for measured values
of \tHeII. This $\eta_{\, \rm thin}$ along with the 
measurements of $\Gamma_{\rm HI}$ provides $\Gamma_{\rm He II}$ 
(using Eq.~\ref{eta_thin_approx}). Here, we estimate $\eta_{\, \rm thin}$ 
using recent measurements of \tHeII~from \citet{Worseck16}. Then we
calculate $\Gamma_{\rm HeII}$ using this $\eta_{\, \rm thin}$  and 
the $\Gamma_{\rm HI}$ measurements from \citet{Becker13}.

In the left-hand panel of Fig.~\ref{fig1} we show \tHeII~measurements of \citet{Worseck16} 
which are calculated at redshift bin intervals of size 0.04 from HST-COS observations of 17
QSO sightlines having He~{\sc ii} Lyman-$\alpha$ forest. 
We calculate $\eta_{\, \rm thin}$ corresponding to each of these 
\tHeII~measurements. These are shown in the right-hand panel of Fig.~\ref{fig1}. The 
error-bars on $\eta_{\rm thin}$ arise from 1-$\sigma$ errors on \tHeII. We need
$\eta_{\rm thin}$ to have values in the range of 40 to 320 to reproduce the observed 
distribution of \tHeII. Note that, the $\eta_{\, \rm thin}$ calculated in this way ignores 
the differences in the \tHI~one expects for different line-of sights. Although, 
the line-of-sight average \tHI~at the regions where \tHeII~was measured
show very good agreement with the mean \tHI~\citep[][ the same mean 
\tHI~that has been used to obtain $f(N_{\rm HI},z)$
by \citealt{InoueAK14}]{FG08, Becker13Ly}, significant variations in 
\tHI~occur on the $\Delta z=0.04$ scales \citep[see figure 8 of][]{Worseck16}.

To estimate $\Gamma_{\rm HeII}$, we need $\eta_{\, \rm thin}$ value in the same
redshift range as the $\Gamma_{\rm HI}$ measurement. Therefore,   
we take median of the \tHeII~measurements in three redshift bins that are  
$z=2.3-2.6$, $z=2.6-3.0$ and $z=3.0-3.5$. These bins match closely with the 
redshift bins used for $\Gamma_{\rm HI}$ measurements by \citet{Becker13}. 
Here, instead of using mean redshift for bins, we use the median redshift 
since the distribution of \tHeII~in each bin is not uniform. 
The median \tHeII~values in these bins are shown in the left-hand panel
of Fig.~\ref{fig1} and provided in the Table~\ref{t1}. 
The error-bars are the 95th percentile values of the distribution of 
errors in each bin. Since, the highest redshift-bin 
contains most of the lower limits on \tHeII~measurements, the median \tHeII~in 
this bin is also a lower limit. The $\eta_{\, \rm thin}$  values 
required to obtain these binned \tHeII~measurements are shown in the
right-hand panel of Fig.~\ref{fig1} and provided in Table~\ref{t1}. 
Error-bars on $\eta_{\, \rm thin}$ are obtained from the error-bars on 
median \tHeII~as shown in the left-hand panel of Fig.~\ref{fig1}.
The median \tHeII~and $\eta_{\, \rm thin}$ show clear increasing trend 
with redshift. We obtain $\Gamma_{\rm He II}$ for these $\eta_{\, \rm thin}$ 
values (from Eq.~\ref{eta_thin_approx}) using the  $\Gamma_{\rm HI}$ measurements of \citet{Becker13}
in the corresponding redshift bins. Table~\ref{t1} summarizes our 
estimated $\Gamma_{\rm He II}$ values as well as the $\Gamma_{\rm HI}$ 
measurements that are used for obtaining them. The errors on $\Gamma_{\rm He II}$ 
also account for the errors on $\Gamma_{\rm HI}$ measurements.
Note that the $\Gamma_{\rm HeII}$ calculated in this way depends 
only on the $f(N_{\rm HI}, z )$ and does not depend on the UVB models. 
Our  $\Gamma_{\rm HeII}$  values are consistent with the values 
obtained by \citet{Worseck16} using their semi-analytic model for 
post-reionization \tHeII. We have also calculated the mean free path 
for He~{\sc ii} ionizing photons \citep[$\lambda_{\rm mfp}$; using 
Eq. 12 and 13 from][]{Khaire13} that depends on 
$\eta$ and $f(N_{\rm HI}, z )$, as given in Table~\ref{t1} in units of proper Mpc. 
Errors on $\lambda_{\rm mfp}$ correspond to errors on the $\eta_{\rm thin}$
values. 

In the next section, we discuss the implications of these inferred $\Gamma_{\rm HeII}$ 
and \tHeII~measurements for calculations of the UVB. 

\section{Helium ionizing UVB}\label{sec3}
We are interested in computing the He~{\sc ii} ionizing UVB to obtain 
$\Gamma_{\rm HeII}$ and \tHeII. This will be used in comparison 
with the \tHeII~measurements and the He~{\sc ii} reionization history 
to constrain the He~{\sc ii} ionizing QSO emissivity. In this section, 
we explain the basic theory to calculate the He~{\sc ii} ionizing 
UVB, the assumptions involved in estimating He~{\sc ii} ionizing emissivity and
theory for calculating He~{\sc ii} reionization history.

\subsection{The UVB}
%
\begin{figure*}
\centering
\includegraphics[totalheight=0.37\textheight, trim=0.3cm 0.3cm 2.0cm 0cm, clip=true]{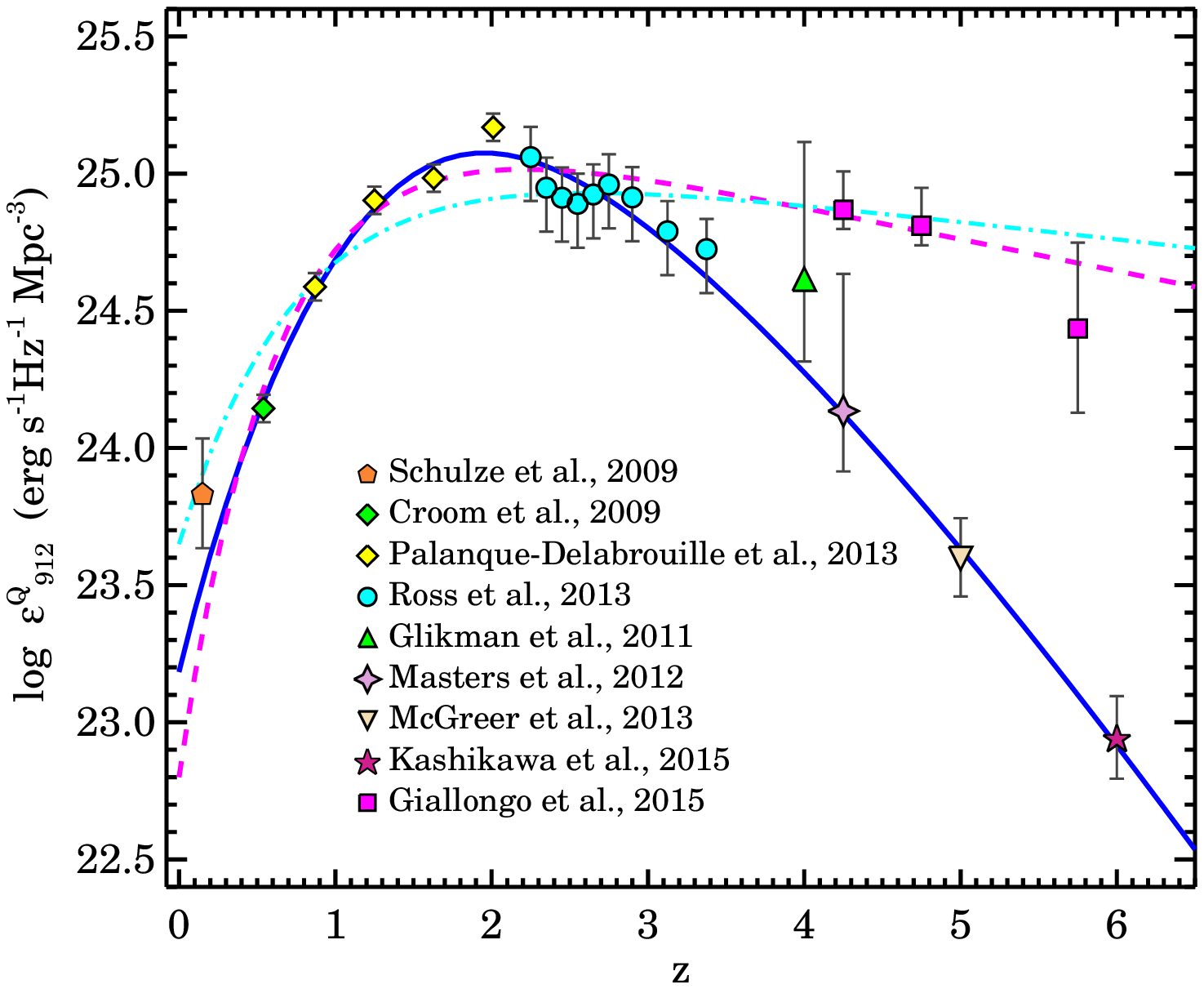}
\caption{The QSO emissivity at $912$\AA~($\epsilon^Q_{\nu_{912}}$) with $z$. Data points
are taken from the compilations of \citet[][see their Table 1]{Khaire15puc} which used recent luminosity function
of optically selected QSOs \citep{Schulze09, Croom09, Glikman11, Masters12, Ross13, Palanque13, McGreer13, Kashikawa15} 
and the emissivity from QSO luminosity function by \citet{Giallongo15}. Blue solid curve is a simple fit through the 
$\epsilon^Q_{\nu_{912}}$ obtained using optically selected QSOs (see Eq.~\ref{eqso}; model A). 
Magenta dashed \citep[model B1 from Eq.~\ref{eqso_high};][]{Khaire16} and cyan dot-dashed curve 
(model B2 from Eq.~\ref{eqso_high_m15}; \citetalias{Madau15}) are 
fits that include $\epsilon^Q_{\nu_{912}}$ from \citet{Giallongo15} at $z>4$. 
}
\label{fig2}
\end{figure*}
%
%
The photoionization rate, $\Gamma_x(z)$, at redshift $z$ for species $x$ 
is obtained by following integral,
%
\begin{equation}\label{gama}
\Gamma_{x}(z)=\int_{\nu_{x}}^{\infty}d\nu\,
\frac{4\pi\,J_{\nu}(z)}{h\nu}\,\sigma_{x}(\nu)\,\,.
\end{equation}
%
Here, $\nu_{x}$ ans $\sigma_x$ are the ionization threshold frequency and 
photoionization cross-section for the species $x$, respectively, $h$ is 
Planck constant and $J_{\nu}(z)$, in units of ergs cm$^{-2}$ 
s$^{-1}$ Hz$^{-1}$ sr$^{-1}$, is the angle averaged specific intensity of 
the UVB radiation at frequency $\nu$ and redshift $z$. $J_{\nu_0}(z_0)$ is 
obtained by solving following cosmological radiative transfer equation 
\citep[see][]{Peebles93, HM96},
%
\begin{equation}\label{Eq.uvb}
J_{\nu_{0}}(z_{0})=\frac{c}{4\pi}\int_{z_{0}}^{\infty}\,dz\frac{(1+z_{0})^{3}\,
\epsilon_{\nu}(z)}{(1+z)H(z)} e^{-\tau_{\rm eff}(\nu_{0},z_{0},z)}.
\end{equation}
%
Here, $c$ is the speed of light, 
$H(z)=H_0 \sqrt{\Omega_m(1+z)^3+\Omega_{\Lambda}}$ is the Hubble parameter, 
frequency $\nu$ is related to $\nu_0$ by $\nu=\nu_0(1+z)/(1+z_0)$, 
and $\epsilon_{\nu}(z)$ is the comoving emissivity of the sources. 
$\tau_{\rm eff}(\nu_{0}, z_{0}, z)$ is an effective optical depth encountered by 
a photon observed at $z_0$ having frequency $\nu_0$ while traveling from its 
emission redshift $z$ to $z_0$. 
Assuming that the IGM clouds along any line-of-sight are Poisson 
distributed, the $\tau_{\rm eff}$ is given by \citep[see][]{Paresce, Paddy3}, 
%
\begin{equation}\label{taueff}
\tau_{\rm eff}(\nu_{0}, z_{0}, z)=\int_{z_{0}}^{z}dz'\int_{N^{\rm min}_{\rm HI}}^{\infty}
dN_{\rm HI}f(N_{\rm HI}, z') (1-e^{-\tau_{\nu'}}).
\end{equation}
%
Here, $ \tau_{\nu'}$ is the continuum optical depth encountered by photons
emitted at frequency $\nu'$ while traveling from their emission redshift 
$z'$ to $z_0$. It is given by, 
%
\begin{equation}\label{tauc_full}
\tau_{\nu'}=N_{\rm HI}{\sigma_{\rm HI}(\nu')}+N_{\rm HeI}{\sigma_{\rm HeI}(\nu')} 
+N_{\rm HeII}{\sigma_{\rm HeII}(\nu')},
\end{equation}
%
where, $\nu'=\nu_{0}(1+z')/(1+z_{0})$. In the redshift range of our interest ($z<4$) 
He~{\sc i} has negligible contribution to $\tau_{\nu'}$ \citep[see also][]{FG09, HM12}. 
Therefore, we approximate $\tau_{\nu'}$ as, 
%
\begin{equation}\label{tauc}
\tau_{\nu'}=N_{\rm HI}[\,\sigma_{\rm HI}(\nu')+\eta \,\sigma_{\rm HeII}(\nu')\,]\,.
\end{equation}
%
Note that, here the $\tau_{\rm eff}$ depends on $\eta(N_{\rm HI})$ and not
just on $\eta_{\rm  \, thin}$. The UVB is obtained by iteratively solving 
Eq.~\ref{quad}-\ref{tauc} for an assumed ionizing emissivity $\epsilon_{\nu}(z)$. 

Here, we are interested in calculating the He~{\sc ii} ionizing UVB at $2<z<4$. 
For that, we need He~{\sc ii} ionizing emissivity (at $\lambda \le 228$\AA) 
and $\Gamma_{\rm HI}$ to estimate $\eta$. Since, we are using the measured values 
of $\Gamma_{\rm HI}$ at $z>2$, we do not need to explicitly calculate the H~{\sc i}
ionizing UVB. However, note that, to calculate the He~{\sc ii} ionizing UVB 
at $z=z_0$ we need $\Gamma_{\rm HI}(z)$ at $z>z_0$. Therefore, in our UVB calculations, 
along with the $\Gamma_{\rm HI}$ measurements by \citet{Becker13} at  $2.4 \le z \le 4.8$, we use
$\Gamma_{\rm HI}$ at $z=2$ from \cite{Bolton07} and at $z>5$ from  \citet{Calverley11} and 
\citet{Wyithe11}. We also estimate the UVB for 1-$\sigma$ higher 
and lower values of measured $\Gamma_{\rm HI}(z)$ to study the 
uncertainties arising in our results due to the uncertainties in the measured  $\Gamma_{\rm HI}$. 

The following subsection explains the usual procedure to estimate the He~{\sc ii}
ionizing emissivity.
%
\subsection{Helium ionizing emissivity}
In the absence of population-{\sc iii} stars at the redshifts of our interest, 
star-forming galaxies emit a negligible amount of He~{\sc ii} ionizing photons.
Therefore, the helium ionizing emissivity $\epsilon_{\nu}$ at 
$\lambda \leq 228$\AA~is contributed by QSOs alone. 
Using the expression for QSO emissivity at 912\AA~($\epsilon^Q_{912}$) and 
the mean SED of QSOs at $\lambda \le 912$\AA~which is usually approximated 
as a power-law $f_{\nu} \propto \nu^{\alpha}$, the $\epsilon_{\nu}$ can be written as, 
%
\begin{equation}\label{Eq.eps}
\epsilon_{\nu}(z) = \Big(\frac{\nu}{\nu_{912}}\Big)^{\alpha} \epsilon^Q_{\nu_{912}}(z), 
\end{equation} 
%
where, $\nu_{912}=c/912$\AA~Hz.

Helium ionizing emissivity depends on $\epsilon^Q_{\nu_{912}}$ and $\alpha$. 
The $\epsilon^Q_{\nu_{912}}$ is obtained from QSO 
luminosity function along with the mean SED from optical to extreme UV wavelengths 
(up to $\sim 912$\AA) that is well observed. However, at $\lambda \leq 912$\AA, 
the power-law index $\alpha$ is measured only 
up to $\lambda \sim 425$\AA~(see Table~\ref{t2}). In absence of any observational 
constraints, this emissivity is usually extrapolated to smaller wavelengths 
(up to $\sim 25$\AA) to estimate the He~{\sc ii} ionizing emissivity. 
Moreover, the values of $\alpha$ reported in the literature
over last two decades are not consistent with each other. Reported values vary from 
-0.56 to -1.96 as summarized in the Table~\ref{t2}. 
The estimates of He~{\sc ii} ionizing UVB and the $\Gamma_{\rm HeII}$ are severely affected 
by the choice of $\alpha$ in the UVB models. These issues motivate us to constrain the $\alpha$ at 
$\lambda \leq 228$\AA~that is consistent with \tHeII~measurements and $\Gamma_{\rm HeII}$.
 For that, we use two models of $\epsilon^Q_{\nu_{912}}(z)$, namely 
model A and model B, as explained below: 

\begin{itemize}
\item {\bf Model A}: 
The model A uses the QSO luminosity functions observed 
at UV and optical wavebands at all redshifts as compiled in \citet[][see their Table 1]{Khaire15puc}.
To estimate the He~{\sc ii} ionizing emissivity and UVB, model A takes $\alpha$ as a free parameter and
$\epsilon^Q_{\nu_{912}}(z)$ in units of  ${\rm erg\, s^{-1}\, Hz^{-1} Mpc^{-3}}$ as \citep{Khaire15puc},
%
\begin{equation}\label{eqso}
\epsilon^Q_{\nu_{912}}(z)=10^{24.6}\,(1+z)^{5.9}\,\frac{\exp(-0.36z)}{\exp(2.2z) + 25.1}\,\,.
\end{equation}
%
This is a simple fit through the compiled $\epsilon^Q_{\nu_{912}}$ values as shown in Fig.~\ref{fig2} (blue solid curve). 
This model needs additional contribution to H~{\sc i} ionizing photons from star-forming galaxies to reionize H~{\sc i} at $z>5.5$ 
and to be consistent with the $\Gamma_{\rm HI}$ measurements at $z>3$ \citep[see][]{Khaire16}.

\item {\bf Model B}: 
In addition to the QSO luminosity functions observed at UV and optical wavebands at $z<4$, model B uses
the QSO luminosity function from \citet{Giallongo15} at $z>4$ obtained by selecting QSO candidates based on their X-ray fluxes.
In contrast with model A, model B do not require any contribution from star forming galaxies to reionize H~{\sc i} i.e QSOs alone
reionize H~{\sc i} in this model \citep[for e.g.,][ hereafter MH15]{Khaire16, Madau15}. Therefore, the H~{\sc i} ionizing emissivity obtained 
through choice of $\alpha$ and $\epsilon^Q_{\nu_{912}}(z)$ in model B has to simultaneously satisfy the 
observational constraints on H~{\sc i} reionization \citep{Planck16, Schenker14, McGreer15} at $z>5.5$, unresolved 
X-ray background at $z>5$ \citep{Moretti12} and $\epsilon^Q_{\nu_{912}}$ obtained by \citet{Giallongo15} at $z>4$.
These constraints provide little room to change $\alpha$ for a given $\epsilon^Q_{\nu_{912}}(z)$ in model B. It is 
unlike the model A where the discrepancy in H~{\sc i} ionizing photons due to decreasing value of $\alpha$ 
can be resolved by increasing the contribution from star-forming galaxies. 
Therefore, instead of making $\alpha$ as a free parameter, for fixed value of $\alpha$ and corresponding $\epsilon^Q_{\nu_{912}}(z)$ 
we explore a break in QSO SED at He~{\sc ii} ionizing part (${\rm E} \ge 4$ Ryd) required to satisfy the \tHeII~measurements.
In model B, we take two values of $\alpha$ and the corresponding two forms of $\epsilon^Q_{\nu_{912}}(z)$ 
that are shown to be consistent with the constraints mentioned above.
First, we take $\alpha=-1.4$ \citep[consistent with][]{Stevans14} and $\epsilon^Q_{\nu_{912}}(z)$ as 
%
\begin{equation}\label{eqso_high}
\log \epsilon^Q_{\nu_{912}}(z)=25.35\exp(-0.0047z)-2.55\exp(-1.61z).
\end{equation}
%
This is consistent with the model presented in \citet{Khaire16}. We denote this combination of
$\alpha$ and $\epsilon^Q_{\nu_{912}}(z)$ as model B1. Second, we take
$\alpha=-1.7$ \citep[consistent with][]{Lusso15} and $\epsilon^Q_{\nu_{912}} (z)$ as
%
\begin{equation}\label{eqso_high_m15}
\log \epsilon^Q_{\nu_{912}}(z)=25.15\exp(-0.0026z)-1.5\exp(-1.3z).
\end{equation}
%
This is the model presented in \citetalias{Madau15}. We denote this combination of
$\alpha$ and $\epsilon^Q_{\nu_{912}}(z)$ as model B2.
We show both of them along with the compiled data in Fig.~\ref{fig2}.
\end{itemize}

Note that, while calculating the He~{\sc ii} ionizing UVB, we also take into account 
the emissivity from diffuse He~{\sc ii} Lyman continuum emission by following 
the prescription given in \citet{HM12} and \citet{FG09}. 
He~{\sc ii} ionizing emissivity is important to calculate the He~{\sc ii} reionization
history. For each of the model emissivities mentioned above, we also estimate the
He~{\sc ii} reionization history following the standard prescription as 
mentioned in the next subsection.
%
%
%
%
\begin{figure*}
\centering
\includegraphics[totalheight=0.75\textheight, trim=0.0cm 0.8cm 7.5cm 0.0cm, clip=true, angle=270]{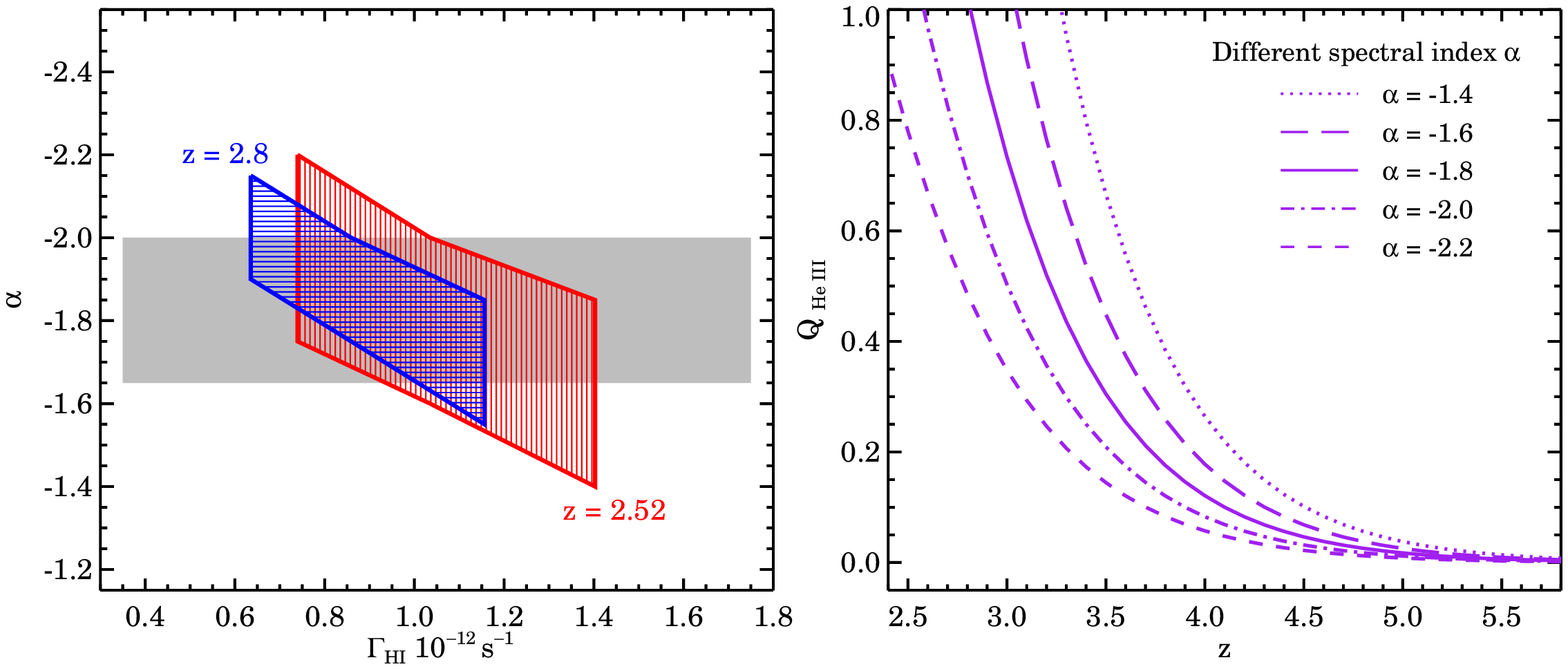}
\caption{Left-hand panel: Joint constraints on the values of power-law index $\alpha$ of mean QSO SED 
(at $\lambda \leq 912$\AA) and $\Gamma_{\rm HI}$ in order to satisfy the binned 
\tHeII~measurement as given in Table~\ref{t1}.  These are obtained for $\epsilon^Q_{\rm 912}(z)$ 
from model A (Eq.~\ref{eqso}). Vertical striped region shows result for 
lowest redshift bin with median $z=2.52$ and horizontal striped region shows results for next redshift 
bin with median $z=2.8$. We do not perform such analysis for highest-$z$ bin where \tHeII~is a lower limit. 
The gray shaded region show the range in $\alpha$ consistent with the redshift of He~{\sc ii} reionization 
$2.6<z_{\rm re}<3.0$. Right-hand panel: $Q_{\rm He III}(z)$ obtained for model A with different $\alpha$. 
}
\label{fig3}
\end{figure*}
\subsection{Helium reionization}\label{sec3.3}
We calculate reionization history of He~{\sc ii} by solving 
following differential equation to estimate the volume averaged
He~{\sc iii} fraction \citep[$Q_{\rm HeIII}$;][]{Shapiro87, Madau99, Barkana01}
%
\begin{equation}\label{dqdt}
\frac{dQ_{\rm HeIII}}{dt}= \frac{\dot n(t)}{\langle n_{\rm He} \rangle}-
\frac{\alpha^{\rm B}_{\rm He II}({\rm T})\,\chi \,C \langle n_{\rm He} \rangle Q_{\rm HeIII}}{a^3(t)} \,\,.
\end{equation} 
%
Here, $\langle n_{\rm He} \rangle=1.87\times10^{-7} y_p/(4-4y_p)$ cm$^{-3}$ is 
the comoving number density of helium, $\dot n(t)$ is comoving number 
density of He~{\sc ii} ionizing photons per unit time, C is the clumping 
factor of He~{\sc ii}, $\chi$ is number of photo-electrons per hydrogen
atom, $a(t)$ is the scale factor and $\alpha^{\rm B}_{\rm HeII}({\rm T})$ 
is the case B recombination coefficient of He~{\sc ii}. 
Here, $\dot n(t)$ is obtained by
%
\begin{equation}\label{ndot}
\dot n(z) = \int^{\infty}_{\nu_{228}}d\nu\,{\Big(\frac{\nu}{\nu_{912}}\Big)^{\alpha}\,\,\frac{\epsilon^Q_{912} (z)}{h \nu}}\,\,,
\end{equation} 
%
where, $\nu_{228}=c/228$\AA~Hz and $\nu_{912}=c/912$\AA~Hz. The solution to the Eq.~(\ref{dqdt}), $Q_{\rm He III}$, at any redshift $z_0$ is given by,
%
\begin{equation}\label{q}
\begin{aligned}
Q_{\rm He III}(z_0)=\frac{1}{\langle n_{\rm He} \rangle} \int^{\infty}_{z_0} dz\,\frac{\dot n(z)}{(1+z)H(z)}\,\times\\ 
\exp \Bigg[-\alpha^{\rm B}_{\rm He II}(T) \langle n_{\rm He} \rangle 
 \int^{z}_{z_0}{dz'\frac{\chi C(z')(1+z')^2}{H(z')}}\Bigg] \,\,.
\end{aligned}
\end{equation} 
%
The process of helium reionization is complete when $Q_{\rm He III}(z_{\rm re})$
becomes unity and that $z_{\rm re}$ is called as reionization redshift.
We take clumping factor from cosmological hydrodynamical simulations of 
\citet{Finlator12} as $C(z)=9.25-7.21 \log(1+z)$. Note that, if instead we use 
$C(z)$ from \citet{Shull12} then the obtained $z_{\rm re}$ for model A is higher
by 0.05. In the He~{\sc iii} regions, we take $\chi=1+[y_p/(2-2y_p)]$ and 
T=20000K to solve for $Q_{\rm He III}(z)$. 
%
%
%
\section{Results and discussion}\label{sec4}
Following the procedure mentioned above, we calculate the He~{\sc ii} 
ionizing UVB and the He~{\sc ii} reionization history for QSO emissivities  from model A and B.  
The results of which are discussed in the following subsections.

\subsection{Model A: constraints on $\alpha$}\label{sec4.1}
The He~{\sc ii} ionizing UVB depends not only on the He~{\sc ii} ionizing emissivity 
from QSOs but also on the $\Gamma_{\rm HI}(z)$ through the calculations of $\eta$. 
The $\Gamma_{\rm HI}(z)$ depends on emissivity from both QSOs and galaxies. Therefore,
the $f_{\rm esc}$ which decides the galaxy contribution to $\Gamma_{\rm HI}$, 
also affects the the He~{\sc ii} ionizing UVB as shown in \citet{Khaire13}. Here, since 
we directly use the measured values of $\Gamma_{\rm HI}$ to calculate the 
He~{\sc ii} ionizing UVB, we do not need to calculate the $f_{\rm esc}$ explicitly. 
We refer reader to \citet{Khaire16} for the required values of $f_{\rm esc}$ 
to obtain the $\Gamma_{\rm HI}$ measurements that are used here.  

We first consider the model A for which the emissivity is obtained 
from QSO luminosity function from UV and optical surveys, as given in Eq.~\ref{eqso}. 
With this emissivity, we calculate the He~{\sc ii} ionizing UVB by varying the spectral index 
$\alpha$\footnote{Note that the $\epsilon^Q_{912}(z)$ given in Eq.\ref{eqso} is obtained for 
$\alpha=-1.4$ at $\lambda \le 1000$\AA. Therefore, when we vary $\alpha$ we multiply 
$\epsilon^Q_{912}(z)$ by a correction factor $k=(1000/912)^{1.4+\alpha}$.}. 
For each $\alpha$ we also vary $\Gamma_{\rm HI}(z)$ within its 1-$\sigma$ uncertainty.  
The calculated UVB for each  $\alpha$ and $\Gamma_{\rm HI}$  provides $\Gamma_{\rm He II}(z)$ and $\eta(z)$. 
Using this $\eta(z)$ in Eq.~\ref{tau_he} and \ref{wn}, we calculate \tHeII($z$). 
In this way, we generate \tHeII($z$) for UVB models with different $\alpha$ and 
$\Gamma_{\rm HI}$. This along with \tHeII~measurements helps us to constrain values of $\alpha$.  
\begin{figure*}
\centering
\includegraphics[totalheight=0.75\textheight, trim=0cm 0.8cm 7.7cm 0.0cm, clip=true, angle=270]{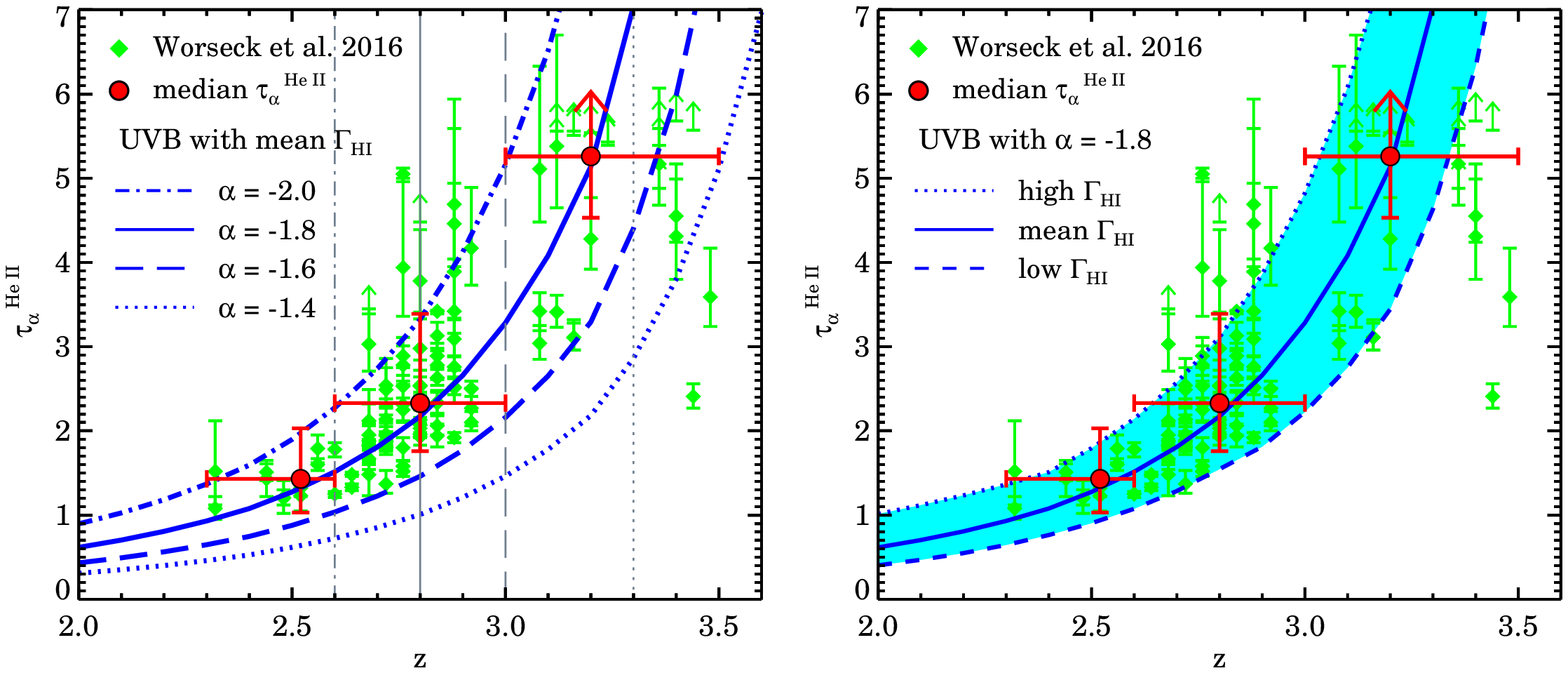}
\caption{Left-hand panel: \tHeII(z) estimated from UVB models obtained for $\epsilon^Q_{\rm 912}(z)$ 
of model A (Eq.~\ref{eqso}) with different 
spectral index $\alpha$ of the mean QSO SED (for $\lambda \leq 912$\AA).
Vertical gray lines with different line-styles mark the redshift of He~{\sc ii} 
reionization (see right-hand panel of Fig.~\ref{fig3}).  The UVB with 
$\alpha=-1.8$ reproduce the binned \tHeII~measurements. Here $\Gamma_{\rm HI}(z)$ values obtained  
for all the models are consistent with the mean values obtained from \citet{Becker13}. 
Right-hand panel: \tHeII(z) estimated from the UVB with $\alpha=-1.8$ 
with different $\Gamma_{\rm HI}$ consistent with the 1$\sigma$ higher and lower 
values. The shaded region show the range in \tHeII(z) due to uncertainty in the 
measured $\Gamma_{\rm HI}$. In both panels \tHeII~measurements by \citet{Worseck16} 
are shown by diamonds and binned  \tHeII~data by circles.
}
\label{fig4}
\end{figure*}

To obtain the binned \tHeII~measurements, as given in Table~\ref{t1}, we calculate 
the required $\alpha$ in the UVB as a function of $\Gamma_{\rm HI}(z)$ within its measured 
uncertainty. The results are shown in the left-hand panel of Fig.~\ref{fig3}. 
Regions with vertical and horizontal stripes provide the joint constraints on 
$\Gamma_{\rm HI}$ and $\alpha$ that is required to obtain the 
binned \tHeII~at $z=2.52$ and $z=2.8$, respectively. Within 1-$\sigma$ range in 
measured $\Gamma_{\rm HI}(z)$, we need UVB with $-2.2<\alpha<-1.4$ at 
$z=2.52$ and  with  $-2.15<\alpha<-1.55$ at $z=2.8$. We do not calculate the required 
$\alpha$ to satisfy \tHeII~at highest redshift bin which is a lower limit. 

The onset of large scatter in \tHeII~measurements seen at $z>2.7$ suggests that the He~{\sc ii} 
reionization has completed at $z\sim2.7$ \citep{Furlanetto10, Shull10, Worseck11, Worseck16}. 
At $z>z_{\rm re}$ the He~{\sc ii} ionizing UVB may not be uniform 
\citep[see][]{Furlanetto09, Davies14}, therefore, 
predicted \tHeII~may not match the measurements. To find $z_{\rm re}$, we 
have also calculated the reionization history. The obtained 
$Q_{\rm HeIII}(z)$ for models with different $\alpha$ is shown in the right-hand panel of 
Fig~\ref{fig3}. The redshift of He~{\sc ii} reionization depends on He~{\sc ii} ionizing 
emissivity and therefore on $\alpha$. The QSO SED becomes flat for higher $\alpha$ 
that gives higher He~{\sc ii} ionizing emissivity. Therefore, higher values 
of $\alpha$ leads to early He~{\sc ii} reionization. If we impose an 
additional constraint on reionization redshift, such as $2.6<z_{\rm re}<3.0$ 
consistent with the trend in \tHeII~data, we need $-2.0<\alpha<-1.65$. The range in 
required $\alpha$ has shown with gray-shade in the left-hand panel of 
Fig~\ref{fig3}. Combining these constraints obtained with the binned 
\tHeII~and the $z_{\rm re}$ together, $\alpha$ can have values from -1.6 to -2.0.

Measurements of $\alpha$ reported in the literature over last two decades are 
summarized in the Table~\ref{t2}. Let us compare the $-1.6>\alpha>-2.0$ 
obtained here with the recent measurements of it. \citet{Lusso15} obtained 
$\alpha=-1.7\pm0.61$ at $z\sim2.4$ using 53 QSOs where the smallest wavelength 
probed by them is 600\AA. \citet{Stevans14} obtained $\alpha=-1.4\pm0.15$ 
at $z<1.5$ using 159 QSOs observed from HST-COS where the smallest wavelength 
probed by them is 475\AA. However, they had fewer than 10 QSOs which probe 
$\lambda <600$\AA. \citet{Tilton16} compiled 11 new QSOs from HST-COS at $1.5<z<2.1$
where the smallest wavelength probed by them is $\lambda\sim425$\AA. 
They combined these with 9 existing QSOs from \citet{Stevans14} and
measured $\alpha=-0.72\pm0.26$ in wavelength range  $450<\lambda<700$\AA.
The $-1.6>\alpha>-2.0$ obtained by us is consistent with the measurements of 
\citet{Lusso15}. It is within 2-$\sigma$ uncertainty from \citet{Stevans14}. 
However, it is 4-$\sigma$ lower than the measurements of \citet{Tilton16}. 
Note that, our inferred value of $\alpha$ is obtained by modeling the UVB at $\lambda \rm \leq 228$\AA~and 
at $2<z<3.5$. Here, we assumed that the QSO SED at $\lambda \leq 912$\AA~follows a single 
power-law and does not change with redshift, same as assumed in other studies. 
The single power-law assumption may not be true since there are no measurements 
that probe SED at $\lambda <400$\AA. \citet{Tilton16} suggested that a simple power-law 
may not be sufficient to explain the QSO SED, even at $\lambda<700$\AA. 
Moreover, the observed QSOs spectra probing $\lambda<500$\AA~are
biased towards most luminous QSOs. Therefore, one expects that 
these measurements can also be biased. Also, the mean QSO SED may have redshift dependence. 
It is important to study such a redshift dependence of $\alpha$ in the direct observations.
%
%
%
\begin{figure*}
\centering
\includegraphics[totalheight=0.75\textheight, trim=0cm 0.4cm 7.7cm 0.0cm, clip=true, angle=270]{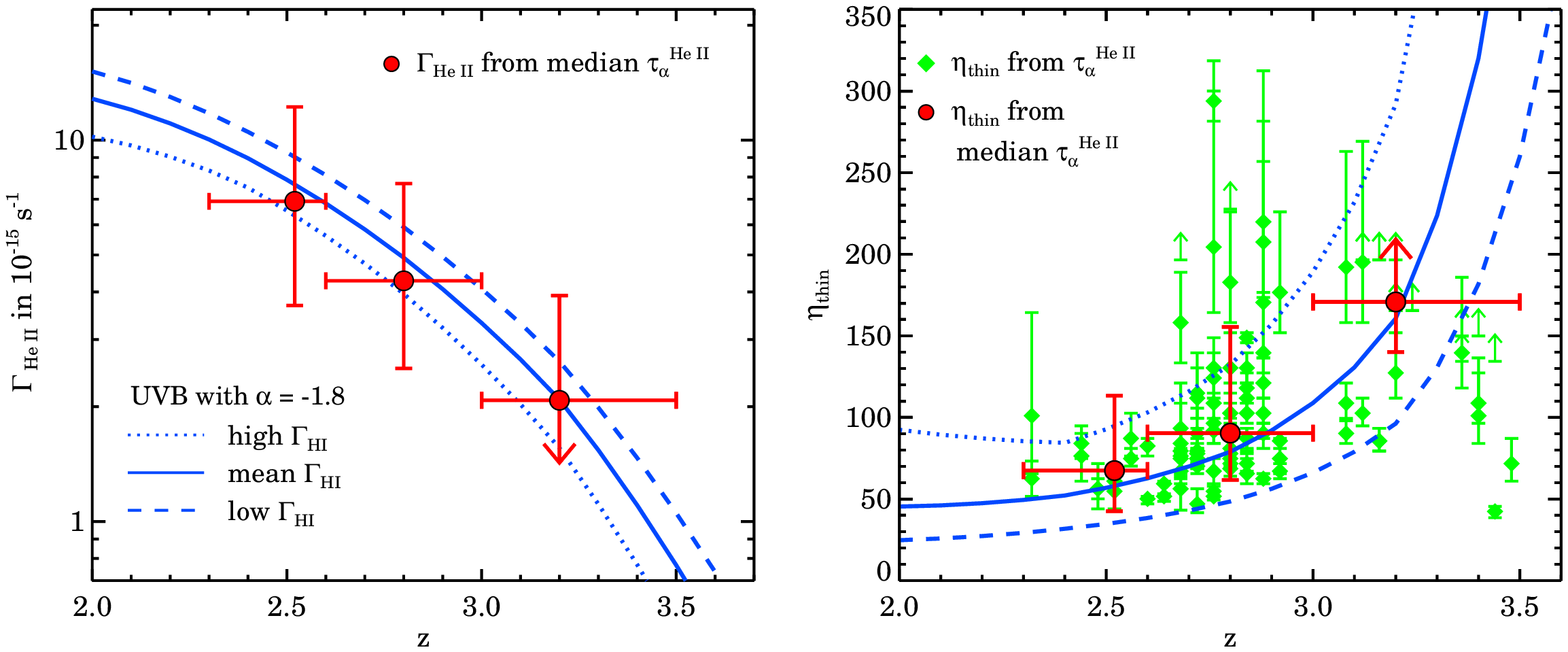}
\caption{$\Gamma_{\rm HeII}(z)$ (\emph{left-hand panel}) and $\eta_{\rm thin}(z)$ (\emph{right-hand panel}) 
obtained from UVB models obtained for $\epsilon^Q_{\rm 912}(z)$ 
of model A (Eq.~\ref{eqso}) with $\alpha=-1.8$. Solid, dash and dotted curves show results obtained from UVB with 
mean $\Gamma_{\rm HI}$, 1-$\sigma$ higher and lower $\Gamma_{\rm HI}$, respectively. Red circles show our estimates of 
$\Gamma_{\rm HeII}$ and  $\eta_{\rm thin}$ from binned \tHeII~data, as described in
Section~\ref{sec2.2} (Table~\ref{t1}).}
\label{fig5}
\end{figure*}
%
%

For the UVB with different $\alpha$ and the mean value of measured 
$\Gamma_{\rm HI}(z)$, the obtained \tHeII($z$) 
is shown in the 
left-hand panel of Fig.~\ref{fig4} along with the measurements from 
\citet{Worseck16} and binned \tHeII~data from Table~\ref{t1}.
It shows that the measured \tHeII~data can be reproduced for 
$-1.6>\alpha >-2.0$. To reproduce binned median \tHeII~data from
Table~\ref{t1}, the UVB with $\alpha=-1.8$ is preferred. 
We also mark the redshift of He~{\sc ii} reionization, $z_{\rm re}$ 
for each $\alpha$. In the post-He~{\sc ii}-reionization era, i.e. 
at $z<z_{\rm re}$, the UVB models are expected to produce the mean 
\tHeII~measurements and may not be at $z>z_{\rm re}$. In the right-hand panel of 
Fig.~\ref{fig4}, we show \tHeII($z$) for the UVB with $\alpha=-1.8$ 
obtained using the mean $\Gamma_{\rm HI}(z)$ as well as 
1-$\sigma$ higher and lower $\Gamma_{\rm HI}(z)$ measurements. 
The shaded region shows the range in \tHeII~arising from the
uncertainty in $\Gamma_{\rm HI}$ measurements. Since it covers most 
of the \tHeII measurements at the post-He~{\sc ii}-reionization era, 
i.e at $z< 2.8$, we prefer the UVB with 
$\alpha=-1.8$. The $\Gamma_{\rm HeII}(z)$ and  $\eta_{\rm thin}(z)$ 
obtained from this UVB are shown in Fig.~\ref{fig5}. Both show good 
agreement with the values estimated from the binned \tHeII~data (from 
Table~\ref{t1}) as explained in Section~\ref{sec2.2}. 
The $\Gamma_{\rm HeII}(z)$ and  $\eta_{\rm thin}(z)$ obtained for 
the UVB with 1-$\sigma$ higher and lower $\Gamma_{\rm HI}(z)$ 
show the spread in these values due to the uncertainty in  
$\Gamma_{\rm H I}(z)$. The very good agreement between the 
$\Gamma_{\rm HeII}(z)$ and  $\eta_{\rm thin}(z)$ obtained 
from the full UVB model and the one estimated using Eq.\ref{tau_he} to 
Eq.~\ref{eta_thin} (see Section~\ref{sec2}), shows the validity of the approximations used in latter.  

All the models mentioned above assume a single power-law SED of 
QSOs at $\lambda \leq 912$\AA. The SED may not be a single-power law; rather it 
can consist of broken power-laws or have breaks at smaller wavelengths.
To obtain the same He~{\sc ii} ionizing emissivity as obtained for our 
preferred model with $\alpha=-1.8$ but with different value of $\alpha$, 
a break in the mean QSO SED at a wavelength 
$228 \le \lambda_b \le 912$\AA~can be applied.\footnote{The purpose of the SED break is to reduce the He~{\sc ii} 
ionizing emissivity. Therefore, it is effective to have at $\lambda_b \ge 228$\AA.} 
The value of the break, the number ($< 1$) 
that is multiplied to the specific intensity at $\lambda \le \lambda_b$, 
can be approximated as $(\lambda_b/912 \rm \AA)^{(1.8+\alpha)}$. For example, 
when we assume $\alpha=-1.4$ consistent with measurements of \citet{Stevans14} 
and \citet{Shull12sed}, we verify that a break in QSO SED at 
$\lambda_b=$228\AA~by a factor of 0.6 gives the same \tHeII($z$) as obtained 
for single power-law SED with $\alpha=-1.8$. Although, the break can be applied at $228 \le \lambda_b \le 912$\AA, 
hereafter we consider the break only at $\lambda_b=228$\AA. 
A slight decrease in the resultant $\Gamma_{\rm HI}$ due to such break in QSO SEDs
can be compensated by marginally increasing $f_{\rm esc}$ from galaxies. 
This SED break can be thought as the escape fraction of He~{\sc ii} ionizing photons from QSOs.  However, in the absence of 
any physical models, such a break in QSO SED and its interpretation should be treated with caution.

We have used $\alpha=-1.4$ in \citet{Khaire16} to estimate the 
required $f_{\rm esc}$  of H~{\sc i} ionizing photons from galaxies 
to obtain the $\Gamma_{\rm HI}$ measurements.
If we use the $\alpha=-1.6$ to $-2.0$ instead, we need an additional 
increase in the predicted $f_{\rm esc}$ in \citet{Khaire16} 
by less than 20 \%.
\subsection{Model B: break in SED}\label{sec4.2}
%
\begin{figure*}
\centering
\includegraphics[totalheight=0.75\textheight, trim=0cm 0.8cm 7.7cm 0.0cm, clip=true, angle=270]{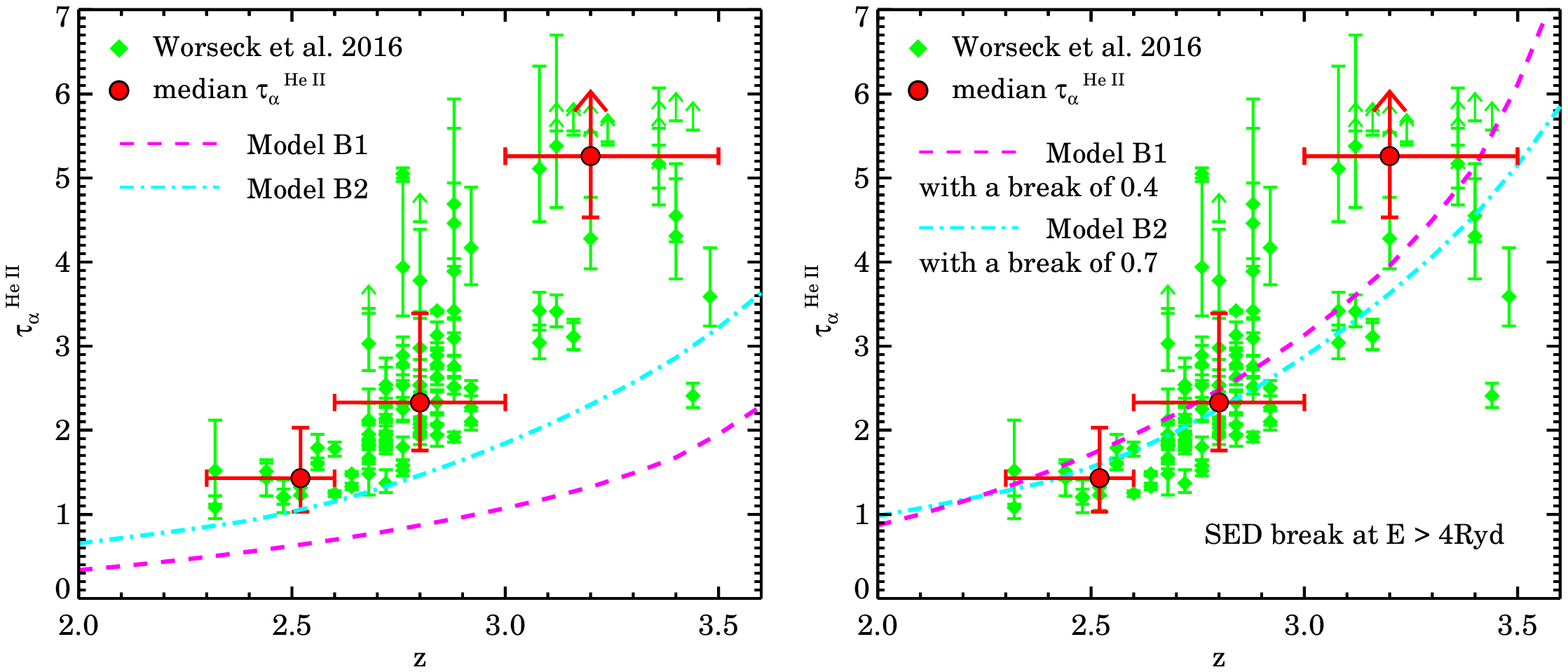}
\caption{Left-hand panel: \tHeII($z$)~obtained from the UVB 
with $\epsilon^Q_{912}(z)$ and $\alpha$ taken from model B1 (Eq.~\ref{eqso_high}; \emph{dashed curve}) and
model B2 (Eq.~\ref{eqso_high_m15}; \emph{dot-dashed curve}). 
Both models fail to reproduce \tHeII~measurements. Right-hand panel: \tHeII~obtained from the UVBs 
with the same $\epsilon^Q_{912}(z)$ and $\alpha$ but with 
appropriate SED breaks at $\lambda \leq 228$\AA~applied 
to match \tHeII~measurements. We need break of $\sim 0.4$ for model B1 (\emph{dashed curve})
and $\sim 0.7$ for model B2 (\emph{dot-dashed curve}). In both panels 
\tHeII~measurements by \citet{Worseck16} are shown by diamonds and binned  
\tHeII~data by circles.
}
\label{fig6}
\end{figure*}
%
%
\begin{figure}
\centering
\includegraphics[totalheight=0.30\textheight, trim=0.8cm 0.4cm 2.2cm 0cm, clip=true]{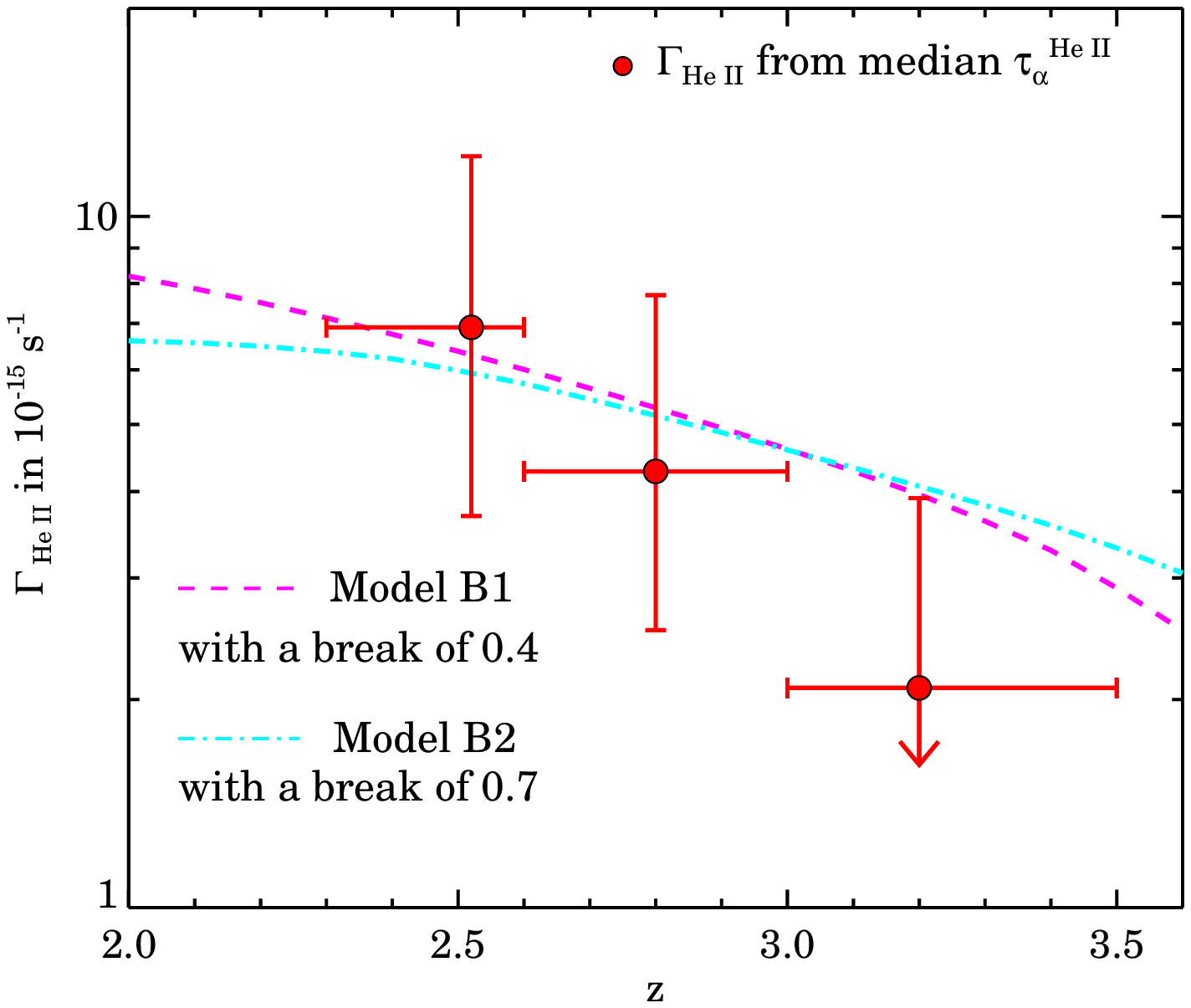}
\caption{$\Gamma_{\rm HeII}(z)$ 
obtained from the UVB models with $\epsilon^Q_{912}(z)$ and $\alpha$
taken from model B1 (Eq.~\ref{eqso_high}; \emph{dashed curve}) and
model B2 (Eq.~\ref{eqso_high_m15}; \emph{dot-dashed curve}) with appropriate SED breaks 
(of 0.4 for model B1 and 0.7 for model B2) 
applied at ${\rm E \ge 4}$ Ryd to match the \tHeII~measurements as shown 
in the right-hand panel of Fig.~\ref{fig6}. The red circles show our estimates 
of $\Gamma_{\rm HeII}$ from binned \tHeII~data, 
as described in Section~\ref{sec2.2} (Table~\ref{t1}).
}
\label{fig7}
\end{figure}
%
%
%
\begin{figure*}
\centering
\includegraphics[totalheight=0.75\textheight, trim=0cm 0.4cm 7.6cm 0.0cm, clip=true, angle=270]{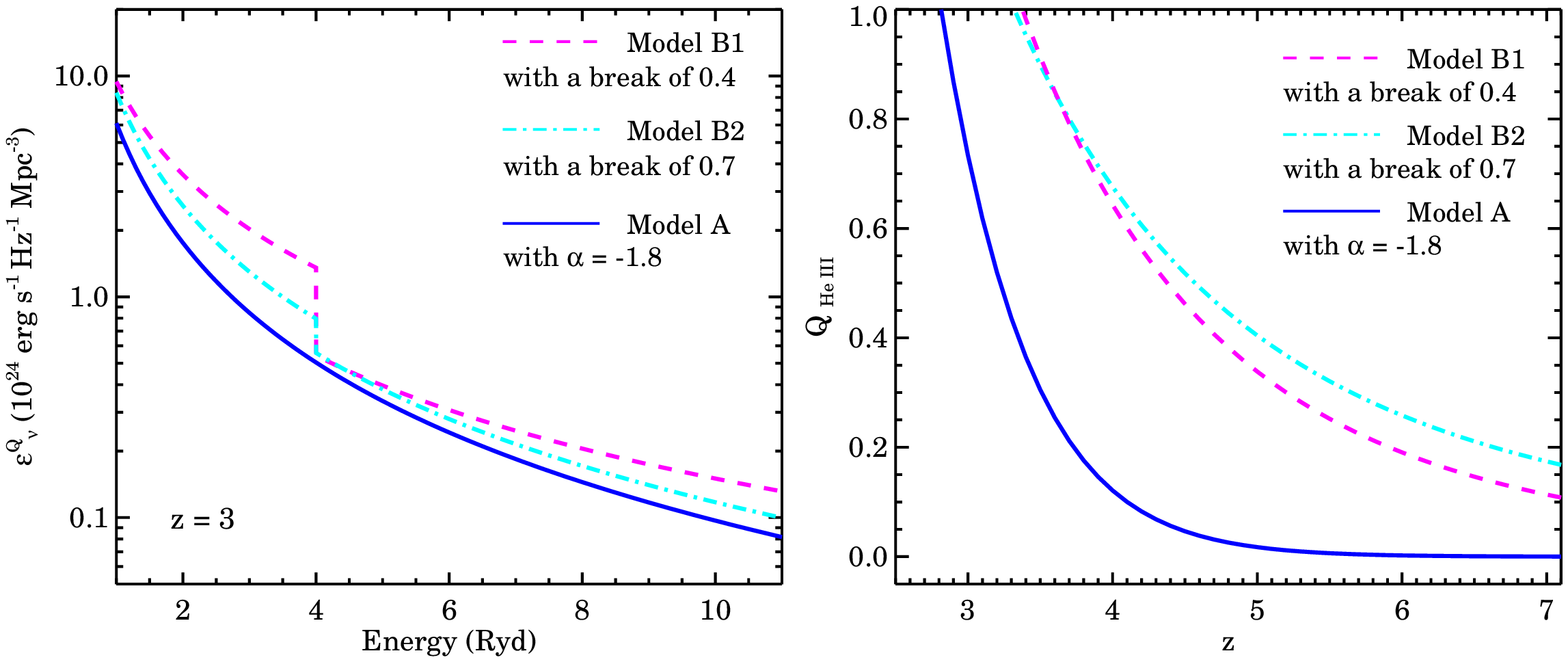}
\caption{The $\epsilon^Q_{\nu}$ at different energies for $z=3$ (\emph{left-hand panel}) and 
$Q_{\rm HeIII}(z)$ (\emph{right-hand panel}) obtained from different models. These models are
model B1 (Eq.~\ref{eqso_high} and $\alpha=-1.4$; \emph{dashed curve}) with SED break of 0.4 at ${\rm E} \ge 4$Ryd, model B2 
(Eq.~\ref{eqso_high_m15} and $\alpha=-1.7$; \emph{dot-dashed curve}) with SED break of 0.7 at ${\rm E} \ge 4$Ryd
and model A (Eq.~\ref{eqso} and $\alpha=-1.8$; \emph{solid curve}) with no break in SED.}
\label{fig8}
\end{figure*}
%

Now we consider the  two combinations of $\alpha$ and $\epsilon^Q_{912}(z)$ 
from model B (Eq.~\ref{eqso_high} and \ref{eqso_high_m15}) that include the emissivity from low-luminosity 
X-ray selected QSOs of \citet{Giallongo15} at $z>4$ and reionize H~{\sc i} 
alone. The model B1 (Eq.~\ref{eqso_high}) uses $\alpha=-1.4$ and 
the model B2 (Eq.~\ref{eqso_high_m15}) uses $\alpha=-1.7$. We calculate the UVB and \tHeII~for these 
models. The results are shown in the left-hand panel of Fig.~\ref{fig6}. The comparison with the 
data shows that these models can not reproduce the \tHeII~measurements. 

These models also predict higher redshift for completion of He~{\sc ii} 
reionization, as $z_{\rm re}=5.2$ for model B1 and  $z_{\rm re}=4.5$ 
for model B2. It is one of the issues of such high QSO 
emissivity models. Therefore, these models need modifications. 
We can not change values of $\alpha$ since they are already adjusted along 
with $\epsilon^Q_{912}(z)$ to reionize H~{\sc i} alone 
without requiring any contribution from galaxies and 
to satisfy different observational constraints on H~{\sc i} reionization. 
However, we can break 
the respective SEDs at $\lambda \leq 228$\AA~so that the H~{\sc i} 
ionizing emissivity and its prediction for H~{\sc i} reionization 
remains the same but the He~{\sc ii} ionizing emissivity reduces.

We estimate the \tHeII~for the UVB obtained with different SED 
breaks at $\lambda \le 228$\AA. We find that for model B1, we need 
SED break of a factor $\sim$0.4 at $\lambda \le 228$\AA~to reproduce the \tHeII~measurements. For 
model B2, since it already has steeper SED with 
$\alpha=-1.7$, a SED break of factor $\sim$0.7 at $\lambda \le 228$\AA~is needed. 
The \tHeII~obtained in these models with such modifications are shown 
in the right-hand panel of the Fig.~\ref{fig6}. The values of 
$\Gamma_{\rm HeII}$ obtained for these models are shown in  Fig.~\ref{fig7}. 
These are in good agreement with the values estimate using binned \tHeII~data. 
In the \emph{left-hand panel} of Fig.~\ref{fig8} we show
the $\epsilon^Q_{\nu}$ at $z=3$ for an illustrative purpose 
from the model B1 and B2 with the SED breaks obtained here. 
For comparison, we also show the $\epsilon^Q_{\nu}$ at $z=3$ from model A with no SED break. 
In all three models, although the H~{\sc i} ionizing emissivities are different, the respective breaks in model B1 and B2
achieve the similar He~{\sc ii} ionizing emissivities as model A.
With such modifications, these models also predict 
lower He~{\sc ii} reionization redshift. For model B1, the $z_{\rm re}$ is 
now 3.4 and for model B2 it is $3.3$. The $Q_{\rm HeIII}(z)$ 
is shown in the right-hand panel of Fig.~\ref{fig8}. Note that if we use the 
clumping factor for He~{\sc ii} from \citet{Shull12} then the obtained 
$z_{\rm re}$ is higher by additional 0.2. The $\epsilon^Q_{912}(z)$ values taken 
in these models are not significantly different from model A 
at $2.3<z<3.2$ (see Fig.~\ref{fig2}). Therefore, the models with SED 
steeper than $\alpha=-1.7$ can be consistent with the 
\tHeII~measurements at $z<3$ but can not reproduce the trend in 
increasing \tHeII~at $z>3$. Also, in such models $\epsilon^Q_{912}(z)$ 
should be higher than the model B2 to reionize 
H~{\sc i} alone that will require higher emissivity than \citet{Giallongo15} 
and it may not be consistent with upper limits on the unresolved X-ray background 
at high-z \citep[see][]{Haardt15}.

The main difference between the model A and model B (both B1 and B2) is the He~{\sc ii} reionization history.
Even though the model B1 and B2 are modified with 
the SED breaks to reproduce the \tHeII~measurements, the $Q_{\rm HeIII}(z)$ 
predicted by them differ significantly from  model A, as shown in the
right-hand panel of Fig.~\ref{fig8}.  For example, at $z\sim4$ (5) in model A 
only 10 (3) per cent of the volume in the Universe is in He~{\sc iii} as compared to the 60 (40) per cent in the model B.
The He~{\sc ii} reionization process is more extended and slower in model B as compared to model A. 
This difference will show imprints on the 
thermal history of the IGM \citep[see also][]{Mitra16, Aloisio16} 
which will be crucial to distinguish these models.  

To distinguish model A where galaxies dominate the H~{\sc i} reionization and model B where QSOs alone reionize H~{\sc i}, 
apart from the thermal history of the IGM the detection of the 21 cm brightness temperature fluctuations will be crucial \citep{Kulkarni17}. 
Also, the independent observational confirmations of the 
QSO luminosity function presented by \citet{Giallongo15} is needed for considering such
high QSO emissivity models. Note that, similar studies such as \citet{Weigel15}, 
\citet{Georgakakis15}, \citet{Ricci17} and \citet{Akiyama17} do not confirm the results of 
\citet{Giallongo15}.

\subsection{Model uncertainties}\label{sec4.3}
Here, we discuss the uncertainties in our models and how they 
affect the results presented in the preceding subsections. 
The estimates of \tHeII~depend on three quantities, the 
assumed $b$-parameter, the $N^{\rm min}_{\rm HI}$ and the $\eta$ 
obtained from the UVB. 

We took $b=28$ km s$^{-1}$ for H~{\sc i} as well as He~{\sc ii} 
assuming that the turbulence dominates the Doppler broadening. If the thermal 
broadening dominates the $b$-parameter then the $b$ for He~{\sc ii} becomes 
14 km s$^{-1}$. This $b$-parameter gives 38\% smaller \tHeII~as compared to the 
one obtained earlier for each UVB model presented here. To match the 
\tHeII~measurements this model will require more steep QSO SED (i.e small $\alpha$)
or small value of break in the QSO SED at $\lambda \le 228$\AA. With this $b$, we find that for QSO emissivity 
from model A, we need $-1.8 > \alpha > -2.0$ to reproduce the 
\tHeII~measurements and to obtain $z_{\rm re} \leq 3$. 
For model B1 and B2, we need a break in QSO SED of 
factor 0.3 and 0.5 at $\lambda \le 228$\AA, respectively, to match 
the \tHeII~measurements. 

The value of $N^{\rm min}_{\rm HI}$  is crucial for 
\tHeII~since the fit to the $f(N_{\rm HI}, z)$ is very steep at low 
values of $N_{\rm HI}$. We took $N^{\rm min}_{\rm HI}$ to have 
minimum equivalent width of $5.2\times10^{-3} {\rm \AA}$, 
which reproduce the \tHI~measurements with $N^{\rm min}_{\rm HI}=10^{12}$ cm$^{-2}$. 
The \tHeII~does not converge rapidly if we extrapolate the fitting form of the 
observed  $f(N_{\rm HI}, z)$ to smaller $N_{\rm HI}$ values. However, note that the 
\citet{InoueAK14} obtained the fit to $f(N_{\rm HI},z)$ at low $N_{\rm HI}$ 
values using the measurements from \citet{Kim13} that probe minimum 
$N_{\rm HI}\sim10^{12.7}$ cm$^{-2}$. For $N_{\rm HI} < 10^{12.5}$ cm$^{-2}$
the $f(N_{\rm HI},z)$ is rather flat and even shows decreasing trend 
\citep[refer to Figure 7 from][]{Odorico16}. If we assume that 
 $f(N_{\rm HI},z)$ is constant or decreasing at $N_{\rm HI} < 10^{12}$ or $10^{12.5}$ cm$^{-2}$
the \tHeII~converges rapidly. When we use a constant $f(N_{\rm HI},z)$ at 
$N_{\rm HI} < 10^{12}$ cm$^{-2}$ and $N^{\rm min}_{\rm HI}=0$, 
we find that the maximum increase in \tHeII~at $z<3.5$ is less than 10\% as compared to
the value we obtain by assuming $N^{\rm min}_{\rm HI}=(16/\eta_{\rm thin})\times 10^{12}\,{\rm cm^{-2}}$ 
and less than 20\% by assuming $N^{\rm min}_{\rm HI}=10^{12}\,{\rm cm^{-2}}$. This does not affect 
our results significantly. 

For the measured values of $\Gamma_{\rm HI}$, values of 
$\eta$ depend on He~{\sc ii} ionizing emissivity. 
We discussed the constraints on the SED, however, we assumed 
fixed $\epsilon^Q_{912}(z)$ values in each model. As mentioned earlier, 
we can not change $\epsilon^Q_{912}(z)$ without changing $\alpha$ in the models that alone reionize H~{\sc i}, 
such as the model B1 and B2. However, we
can change it in the model A. If we uniformly reduce the $\epsilon^Q_{912}(z)$ 
in our model A by 10\% (20\%) at $z>2$ allowed by the uncertainties
in the QSO luminosity functions, we find that the $\eta$ 
increases due to a decrease in He~{\sc ii} ionizing emissivity. This leads to 
higher \tHeII~by $10-15$\% ($25-40$\%) over redshift $2-3.5$. For such models, we find 
that $-1.5 > \alpha > -1.9$ is needed to reproduce the \tHeII~measurements.      

Note that the variation in \tHeII~arising from all these uncertainties 
is smaller than the one arising from the uncertainty in the measured 
$\Gamma_{\rm HI}$ itself (see the right-hand panel of Fig.~\ref{fig4}). 
In future, more stringent constraints on the QSO SED can be obtained 
using accurate measurements of $\Gamma_{\rm HI}$ and more
observations of \tHeII~in the post-He~{\sc ii}-reionization era ($z<2.6$).  
Currently, there are only two sightlines, HE2347$-$4342 and  HS1700$+$6416, that
probe He~{\sc ii} Lyman-$\alpha$ forest at $z<2.6$. 

\section{Summary}\label{sec5}
Here, we present a method that constrains the He~{\sc ii} ionizing emissivity
using \tHeII~measurement obtained from He~{\sc ii} Lyman-$\alpha$ forest and the
distribution of H~{\sc i} in the IGM obtained from H~{\sc i} Lyman-$\alpha$ forest.
The method uses our cosmological radiative transfer code developed to calculate the 
UVB by varying the input He~{\sc ii} ionizing emissivity to be consistent with 
\tHeII~measurements. The He~{\sc ii} ionizing emissivity depends on the QSO 
emissivity obtained from their luminosity functions and the mean QSO 
SED extrapolated at $\rm E \geq 4$ Ryd. The latter
has been observationally constrained only up to $\rm E \sim 2.3$ Ryd. 
We constrain the QSO SED at $\rm E\geq 4$ Ryd required to satisfy
the recent measurements of \tHeII~\citep{Worseck16} using 
models of updated QSO emissivity
at 1 Ryd \citep[][]{Khaire15puc} and H~{\sc i} distribution of the 
IGM \citep{InoueAK14} in our UVB code. We have also calculated the 
$\Gamma_{\rm HeII}$ (provided in Table~\ref{t1}) from the binned 
\tHeII~data which depends only on the H~{\sc i} column density distribution at 
$N_{\rm HI} <10^{16}$ cm$^{-2}$ and the $\Gamma_{\rm HI}$ 
measurements at $z>2.2$ \citep{Becker13}.

The mean SED obtained from QSO composite spectra is usually approximated as 
a power-law
$f_{\rm E} \propto \rm E^{\alpha}$ at $\rm E \geq 1$ Ryd. 
 For QSO emissivity obtained using their luminosity functions from optical surveys, 
we find that the \tHeII~measurements 
are well reproduced when we use the power-law index $-1.6<\alpha<-2.0$. 
The UVB models with this $\alpha$ not only reproduce the majority of the 
\tHeII~measurements but also reionize He~{\sc ii} at $2.6<z_{\rm re} <3.0$,
consistent with the trend seen in the \tHeII~data.
The  $-1.6<\alpha<-2.0$ constrained here is consistent with the 
measurements of \citet{Lusso15} and \citet{Stevans14} but 4-$\sigma$ lower
than the measurement by \citet{Tilton16}. 
We prefer the UVB model with $\alpha=-1.8$ because it reproduces the
\tHeII~measurements and our estimated $\Gamma_{\rm HeII}$ values within the 
uncertainties in the measured $\Gamma_{\rm HI}$.

We also consider models of QSO emissivity that include the luminosity function
obtained from low-luminosity X-ray selected QSOs  presented by \citet{Giallongo15} at $z>4$. 
These models are constructed such that they can reionize H~{\sc i} without requiring 
any contribution from galaxies \citep[\citetalias{Madau15}][]{Khaire16} when extrapolated to $z>6$. 
We find that these models can not reproduce \tHeII~measurements and need 
modifications to reduce the He~{\sc ii} ionizing emissivity. 
For such a model with $\alpha=-1.4$ from \citet{Khaire16}, 
we need a break in mean QSO SED at $\rm E \geq 4$ Ryd of a factor $\sim 0.4$.
Similarly, for a model with $\alpha=-1.7$ from \citetalias{Madau15} we need 
break of a factor $\sim 0.7$ (see the left-hand panel of Fig.~\ref{fig8} for illustration of such SED breaks). 
These modified models give epoch of He~{\sc ii} reionization at $3.3-3.4$
which is significantly smaller than $4.5-5.2$ obtained without 
such modifications. However, even with such modifications 
the He~{\sc ii} reionization history is significantly different from 
standard models (see the right-hand panel of Fig.~\ref{fig8}) which do not include the luminosity function of \citet{Giallongo15}. 
The thermal history of the IGM will play crucial role in distinguishing 
these models.

The method presented here requires better observational constraints on both 
$\Gamma_{\rm HI}$ and the H~{\sc i} distribution in the IGM, as well as measurements of 
\tHeII~over a large redshift range, to accurately constrain the 
mean QSO SED together with its redshift dependence.
Using different QSO SEDs provides significantly different UVB
at He~{\sc ii} ionizing wavelengths. Observations of metal line ratios tracing 
lower and higher energies around He~{\sc ii} ionization potential 
(such as C~{\sc iv} and Si~{\sc iv}) can be considered to test different 
models of the UVB \citep[see for e.g.,][]{Fechner11}. We plan to carry such 
studies in future. 
\section*{acknowledgement} 
VK thanks the anonymous referee for reports that helped to improve
this manuscript. VK also thanks R. Srianand, P. Gaikwad and P. Arumugasamy for useful comments on the manuscript.

\bibliographystyle{mnras}
\bibliography{vikrambib}


\end{document}